\documentclass[reprint, hidelinks, superscriptaddress, amsmath, amssymb, aps, pre]{revtex4-2}

\usepackage{filecontents}

\usepackage{inputenc}
\usepackage[T1]{fontenc}
\usepackage{tcolorbox}
\usepackage{grffile}
\usepackage{fixme}
\usepackage{float}
\usepackage{longtable}
\usepackage{wrapfig}
\usepackage{rotating}
\usepackage[normalem]{ulem}
\usepackage[outline]{contour}
\usepackage{textcomp}
\usepackage{amsmath}
\usepackage{amssymb}
\usepackage{xfrac}
\usepackage{capt-of}
\usepackage{placeins}

\usepackage{algpseudocode}
\usepackage{pgfplots}
\usepackage{pgfplotstable}
\usepackage{relsize}
\usepackage[export]{adjustbox}
\usepackage{lipsum}

\usepackage{bm}
\usepackage{standalone}
\usepackage{tikz}
\newtcolorbox{mybox}[1]{colback=bright_red!5!,colframe=white!0, fonttitle=\bfseries,title=#1}

\usepackage{ifthen}

\usepackage{pgfmath}
\usepackage{pgfplots}
\usepackage{todonotes}

\usepackage{xfp}

\usepackage{xr-hyper}
\usepackage{hyperref}
\usepackage{customcolours}
\usepackage{customcommands}
\usepackage{globalplotconfig}

\usepackage[capitalise]{cleveref}

\pgfplotsset{compat=newest}

\usetikzlibrary{
  shapes.geometric,
  decorations,
  calligraphy,
  positioning,
  calc,
  math,
  3d,
  decorations.pathreplacing,
  decorations,
  patterns,
  decorations.markings,
}
\usepgfplotslibrary{groupplots, fillbetween}
\usepgfplotslibrary{external}
\usetikzlibrary{pgfplots.external}

\begin{document}
\preprint{OPES/123-QED} \title{Particle-Environment Interactions In
  Arbitrary Dimensions: A Unifying Analytic Framework To Model
  Diffusion With Inert Spatial Heterogeneities}

\author{Seeralan Sarvaharman}
 \email{Email: s.sarvaharman@bristol.ac.uk}
\affiliation{Department of Engineering Mathematics, University of Bristol, BS8 1UB, UK}
\author{Luca Giuggioli}%
 \email{Email: Luca.Giuggioli@bristol.ac.uk}
\affiliation{Department of Engineering Mathematics, University of Bristol, BS8 1UB, UK}
\affiliation{Bristol Centre for Complexity Sciences, University of Bristol, BS8 1UB, UK}


\begin{abstract}
  Interactions between randomly moving entities and spatial disorder
  play a crucial role in quantifying the diffusive properties of a
  system. Examples range from molecules advancing along dendritic
  spines, to animal anti-predator displacements due to sparse
  vegetation, through to water vapour sifting across the pores of
  breathable materials. Estimating the effects of disorder on the
  transport characteristics in these and other systems has a long
  history. When the localised interactions are reactive, that is when
  particles may vanish or get irreversibly transformed, the dynamics
  is modelled as a boundary value problem with absorbing
  properties. The analytic advantages offered by such a modelling
  approach has been instrumental to construct a general theory of
  reactive interaction events. The same cannot be said when
  interactions are inert, i.e.~when the environment affects only the
  particle movement dynamics. While various models and techniques to
  study inert processes across biology, ecology and engineering have
  appeared, many studies have been computational and explicit results
  have been limited to one-dimensional domains or symmetric geometries
  in higher dimensions. The shortcoming of these models have been
  highlighted by the recent advances in experimental technologies that
  are capable of detecting minuscule environmental features. In this
  new empirical paradigm the need for a general theory to quantify
  explicitly the effects of spatial heterogeneities on transport
  processes, has become apparent. Here we tackle this challenge by
  developing an analytic framework to model inert particle-environment
  interactions in domains of arbitrary shape and dimensions. We do so
  by using a discrete space formulation whereby the interactions
  between an agent and the environment are modelled as perturbed
  dynamics between lattice sites. We calculate exactly how disorder
  affects movement due to reflecting or partially reflecting
  obstacles, regions of increased or decreased diffusivity, one-way
  gates, open partitions, reversible traps as well as long range
  connections to far away areas. We provide closed form expressions
  for the generating function of the occupation probability of the
  diffusing particle and related transport quantities such as
  first-passage, return and exit probabilities and their respective
  means. The strength of an analytic formulation becomes evident as we
  uncover a surprising property, which we term the disorder
  indifference phenomenon of the mean first-passage time in the
  presence of a permeable barrier in quasi 1D systems. To demonstrate
  the widespread applicability of our formalism, we consider three
  examples that span different spatial and temporal scales. The first
  is an exploration of an enhancement strategy of transdermal drug
  delivery through the stratum corneum of the epidermis. In the second
  example we associate spatial disorder with a decision making process
  of a wandering animal to study thigmotaxis, i.e.~the tendency to
  remain close to the edges of a confining domain. The third example
  illustrates the use of spatial heterogeneities to model inert
  interactions between particles. We exploit this aspect to analyse
  the search of a promoter region on the DNA by transcription factors
  during the early stages of gene transcription.
\end{abstract}

\maketitle
\section{Introduction}\label{sec:intro}
Local interactions between mobile agents or particles and their
environmental features plays a crucial role in the dynamics of many
systems across disciplines and scales~\cite{fischerhertzbook1991,
  amroetal2000, buxtonclarke2006, beyeretal2016, assisetal2019}. When
such environmental features are inert heterogeneities, the local
interactions only affect the movement dynamics of the agents. A wide
array of spatial heterogeneities can be classed as inert,
e.g.~impenetrable or permeable barriers, areas of reduced or increased
mobility, lattice defects such a disclinations, and traps that are
reversible.

In some instances the presence of such heterogeneities is by design,
e.g.~in manufacture engineering where materials are constructed to
have specified diffusive
characteristics~\cite{seebauernoh2010,baietal2018}. In other scenarios
spatial heterogeneities occur naturally. In ecology, animals alter
their foraging behaviour due to variations in vegetation
cover~\cite{jeltschetal2013,silveiraetal2016}. In molecular biology,
particles undergo fence hindered motion in the lipid bilayer membranes
of
eukaryotes~\cite{fujiwaraetal2002,fujiwaraetal2016},
and slow down dramatically when moving within the cell cytoplasm due
to exclusion processes~\cite{mcguffeeelcock2010}. While the
relationship between mobility and spatial disorder in these and other
systems has always been a focus of scientific studies, it is the
highly resolved nature of modern observations that has made apparent
the need for a general framework to model inert particle-environment
interactions.

Investigations on movement dynamics in spatially disordered systems
date back as early as the 50's
\cite{dyson1953,schmidt1957,halperin1965,liebbook1981,horibook1968}.
Despite such a long history most analyses lack a rigorous quantitative
description of the `microscopy' of the interaction events between the
particle and the environment. In the past, transport in highly
disordered media has been studied approximately, linking the
Hausdorff–Besicovitch dimension of fractal structures to a diffusion
constant via scaling arguments \cite{havlinben-avraham1987}. Other
approaches have kept the geometry non-fractal utilising random walks
on regular lattices, the so-called random walk in random environments
model
\cite{zwanzig1982,alexanderetal1981,sinai1983,durrett1986,blumbergselingeretal1989,murthykehr1989,hughesbook1996}. The
majority of these latter studies pertain to 1D domains
\cite{hauskehr1987}, and have used techniques such as the effective
medium approximation to find statistical properties of the movement
dynamics. It is precisely the absence of explicit spatiotemporal
representation of higher dimensional particle dynamics, that has
hampered the widespread applicability of these various models to
current high fidelity observations.

More recent theoretical applications to movement in disordered
environment have focused on the diffusive dynamics in the cell
(e.g. see reviews in
Refs.~\cite{bressloffnewby2013,metzleretal2014a,merozsokolov2015}). Many
attempts in this area are macroscopic and tackle particle dynamics
without representing the local interactions. Some efforts, giving
importance to the very slow dynamics that emerge from overcrowding
effects, have modelled particle movement via fractional diffusion
\cite{hoflingfranosch2013}. The relative size of accessible versus
inaccessible regions has been accounted for using diffusion on
percolation clusters and has highlighted the difference between
compact versus non-compact exploration of space
\cite{meyeretal2012}. Other investigations have put emphasis on the
spatiotemporal dynamics of the environment and have developed the
so-called diffusing-diffusivity models, where the diffusion strength
of the medium itself is a random variable
\cite{chubynskyslater2014,lanoiseleegrebenkov2018,sposinietal2018,lanoiseleeetal2018}.

These approaches have brought important insights and have broadened
the tools and techniques with which to study disordered
systems. However, they too lack the mechanistic connection between the
environmental heterogeneities and the moving particle
\cite{hoflingfranosch2013}. With the advent of new experimental
techniques such as super-resolution microscopy and single particle
tracking \cite{wangetal2012}, the need for an explicit consideration
of particle-environment interactions has also emerged in
microbiology~\cite{metzler2017,lampoetal2017}.

The challenge in fulfilling this need stems from the symmetry breaking
role that disorder plays on the underlying diffusive dynamics. In most
instances describing explicitly multiple heterogeneities is an
unwieldy boundary value problem. The vast majority of theoretical
studies have in fact been limited to highly symmetric scenarios,
e.g.~spherically symmetric domains with concentric layers of different
diffusivity \cite{korabelbarkai2011,
  godecmetzler2015,vaccarioetal2015,godecmetzler2016} and an array of
periodically placed semi-permeable barriers in 1D
\cite{powlesetal1992,kenkreetal2008,kalayetal2008,grebenkov2014,moutalgrebenkov2019}.

To bypass this challenge, and to avoid the use of computationally
prohibitive stochastic simulations, we propose a unifying analytic
framework to model interactions between diffusing agents and spatial
disorder. We do so by developing a random walk theory where
interactions with heterogeneities are represented as a perturbation of
the transition probabilities of a homogeneous lattice. By extending
the so-called defect technique
\cite{montrollpotts1955,montrollwest1979,kenkrebook2021}, we are able
to model explicitly any inert particle-environment interactions in
arbitrary dimensions, e.g. the passage through porous or permeable
barriers, the movement within regions of altered diffusivity, which we
call sticky or slippery sites as well as shortcut jumps to far away
locations.

The theory allows us to derive mathematical expressions for the random
walker occupation probability, the so-called propagator. The
generating function of these propagators are exact and obtained in
terms of the occupation probability in the absence of spatial
heterogeneities, thereby making our framework modular in its
application. Multiple derived quantities, such as first-passage,
return and exit probabilities, which in the past were obtained either
numerically or known only in asymptotic limits
\cite{benichouvoituriez2014}, can now be readily computed via the
evaluation of certain matrix determinants.

Given the generality of our framework, we have opted to provide three
examples of application. The first deals with an extra-cellular
process, namely the potential optimisation of transdermal drug
delivery \cite{prausnitzetal2004,prausnitzlanger2008}. The second
example is the modelling of thigmotaxis, the tendency of insects and
other animals to remain preferentially close to physical boundaries
whilst moving \cite{jeansonetal2003,miramontesetal2014}. The third application
concerns with the search dynamics in a two particle coalescing process
that is of relevance to early stages of gene transcription
\cite{mirnyetal2009,shinkolomeisky2019}.

The remainder of the paper is organised as follows. In
\cref{sec:movement_in_het} we introduce the general mathematical
formalism via a lattice random walk Master equation, and show how we
represent different kinds of heterogeneities. In
\cref{sec:heterogenous_propagator} we solve the Master equation and
find the exact propagator.  \cref{sec:first_passage} deals with
first-passage statistics and their associated mean, i.e. mean
first-passage, mean exit and mean return times.  The latter half of
the paper, Sections~\ref{sec:transdermal}, \ref{sec:thigmotaxis} and
\ref{sec:coalescing}, are devoted to the three applications mentioned
previously, which are transdermal drug delivery, thigmotaxis and gene
transcription. Lastly, conclusions and future applications form
\cref{sec:conclusion}.

\section{Movement In Heterogeneous Environments}
\label{sec:movement_in_het}
We start by defining the dynamics of a Markov lattice random walk
on a $d$-dimensional lattice via
\begin{equation}
  \label{eq:homogenous_master_eq}
  \propsymbol(\nvec, t+1) = \sum_{\mvec{m}}\matjump{\nvec}{\mvec{m}}
  \propsymbol(\mvec{m}, t),
\end{equation}
where $\mvec{n}$ is a $d$-dimensional vector and
$\matjump{\nvec}{\mvec{m}}$ is the transition probability from site
$\mvec{m}$ to site $\mvec{n}$ such that
$\sum_{\mvec{m}}\matjump{\mvec{m}}{\nvec} = 1$ for any site $\nvec$ on
the lattice, i.e. with $d=1$, $\mmat{A}$ is a probability conserving
transition matrix, and when $d > 1$, $\mmat{A}$ is actually a
tensor. For convenience in inverting generating functions, as compared
to Laplace inversion, we use a discrete time formulation with the
variable $t$. Changes to a continuous time description is
straightforward \cite{giuggioli2020}, but is omitted here. We refer to
this equation as the homogeneous Master equation and its solution,
given a localised initial condition, as the {\em homogeneous
  propagator}. The underlying lattice is referred to as the
homogeneous lattice whose size can be finite or infinite.

\begin{figure}[ht]
  \centering
  \includegraphics[width=\columnwidth]{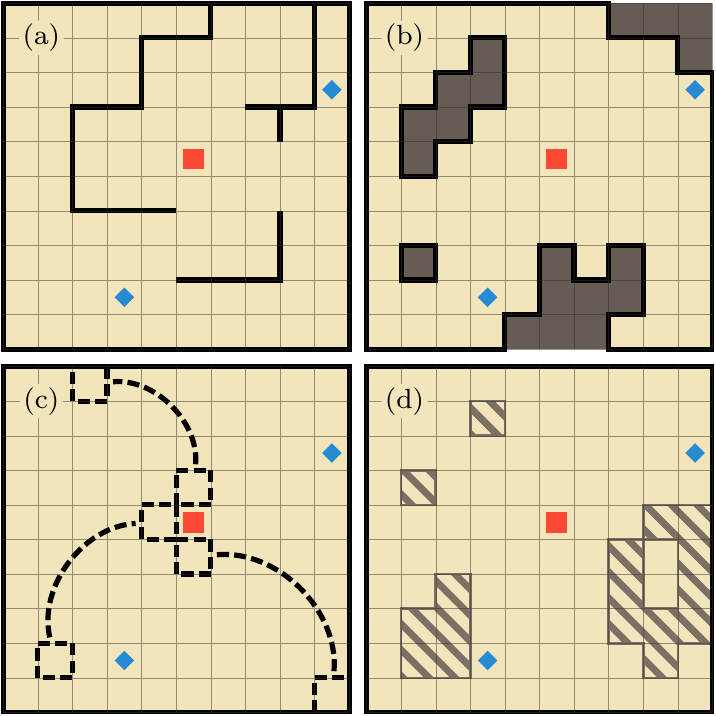}
  \caption{Examples of the spatial heterogeneities within a square
    lattice of width 10 with reflecting boundaries. Panel (a)
    represents an open partitions with the solid black lines
    indicating impenetrable barriers. When these barriers enclose a
    region, some space becomes inaccessible indicated by the sites
    coloured dark grey in panel (b). Panel (c) shows a lattice where
    three pairs of non-neighbouring sites have a long range
    connection, i.e. transitions from a dashed site include the
    nearest-neighbours as well as the site connected via the dashed
    line. Panel (d) is an example of where the diffusivity of the
    striped sites is smaller than the regular (non-striped) sites. The
    central site flagged by a red square and the two sites flagged by
    a blue diamond are, respectively, the initial condition and the
    absorbing targets for use in later sections.}
\label{fig:refl_schematic}
\end{figure}

Since spatially heterogeneous dynamics are defined relative to the
homogeneous system, we define heterogeneities as locations or {\em
  defects} where the dynamics are different from the corresponding
ones on the homogeneous lattice. Examples of heterogeneities are
depicted in \cref{fig:refl_schematic}.

\begin{figure}[ht]
  \centering
  \includegraphics{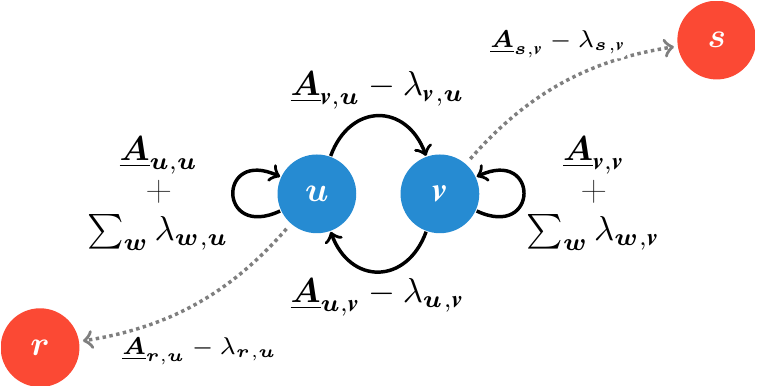}
  \caption{A schematic representation of the transition probabilities
    after the introduction of spatial heterogeneity or disorder. The
    probability of hopping from site $\vs$ to $\vps$ is given by
    $\matjump{\vps}{\vs}$. When $\rjq$ is positive, the probability of
    jumping from $\vs$ to $\vps$ decreases, while the probability of
    staying put increases. When $\rjq$ is negative, the opposite
    effect occurs with a decrease in the probability of staying, while
    increasing the jump probability from $\vs$ to $\vps$. The
    parameter $\ljq$ affects the transition probability from $\vps$ to
    $\vs$ and the probability of remaining at $\vps$ in an equivalent
    manner.}
  \label{fig:defect_schematic}
\end{figure}

The heterogeneities displayed in \cref{fig:refl_schematic} emerge from
the modification of the outgoing transitions from one or more sites,
hence we refer to these altered transitions as heterogeneous
connections.  For example, given a partially reflecting barrier in
between two neighbouring sites, the jump probability from either of
the two sites to the other is reduced, while the probability of
staying put at either of the sites is increased. Conversely, by
connecting together two non-neighbouring sites, we may wish to reduce
the probability of staying put at a given site, whilst adding the
possibility of hopping to the site further away. One can represent
conveniently these or any other heterogeneity through a modification
of the transitions as depicted in
\cref{fig:defect_schematic}. Formally, the outgoing connections of the
sites $\vs$ and $\vps$ are adjusted by introducing the parameters
$\rjq$ and $\ljq$ to create two heterogeneous connections. Although we
choose to modify transitions in both direction, i.e. from $\vs$ to
$\vps$ and $\vps$ to $\vs$, this does not have to be the
case. Modifications of only outgoing connections are also permitted,
e.g. see the dashed arrows connecting the additional sites $\mvec{r}$
and $\mvec{s}$.

The construction implicitly conserves probability, which can be
evinced by picking a defect, e.g.  $\vs$, and summing over all of the
outgoing probabilities. The changes induced by the $\jqsymbol$
parameters cancel each other out leaving
$\sum_{\ws} \matjump{\ws}{\vs}$, with $\ws$ representing all the
neighbours of $\vs$, equal to the homogeneous outgoing probability. To
ensure positive probability for a given heterogeneous site $\vs$, we
have the conditions
\begin{equation}
\rjq[][\ws] \leq \matjump{\ws}{\vs},
\label{eq:lam_const_1}
\end{equation}
for all $\ws$ with a heterogeneous connection in the direction
$\vs$ to $\ws$, and
\begin{equation}
0 \leq \matjump{\vs}{\vs} + \sum_{\ws} \rjq[][\ws],
\label{eq:lam_const_2}
\end{equation}
which enforces upper and lower bounds on the $\jqsymbol$ parameters
although each one of them can be positive or negative. This
formulation allows one to perturb arbitrarily the homogeneous lattice
creating any type of probability conserving particle-environment
interactions.

\subsection{Quantitative representation of heterogeneities}
To understand the practicality of the formalism, we focus on the three
specific types of heterogeneities in \cref{fig:refl_schematic},
namely, barriers (Figs.~(\ref{fig:refl_schematic})a and
(\ref{fig:refl_schematic})b), long-range connections
(\cref{fig:refl_schematic}c) and sticky sites
(\cref{fig:refl_schematic}d). In the following subsection we present
convenient parameterisation for the constant $\jqsymbol$'s to
construct such heterogeneities.

\subsubsection{Barriers and Anti-Barriers}
\begin{figure}[H]
    \centering
    \includegraphics{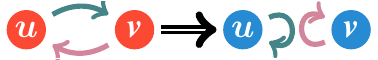}
    \caption{Example of a reflecting barrier between $\vs$ and $\vps$
      generated by modifying the transition probabilities (from the
      left to the right of the schematic). The modified transitions
      are indicated by coloured arrows. The modification in this case
      results in an impenetrable barrier between $\vs$ and $\vps$,
      with $\rdec =\ldec=1$.}
 \label{fig:example_defect_ref}
\end{figure}

With $\vs$ and $\vps$ two neighbouring sites, we construct a partially
reflecting barrier by having $\rjq = \rdec \matjump{\vps}{\vs}$ and
$\ljq = \ldec \matjump{\vs}{\vps}$ where $\rdec, \ldec \in [0, 1]$ is
a measure of the reflectivity of the barrier. When $\rdec, \ldec = 1$
we have an impenetrable barrier (shown in
\cref{fig:example_defect_ref}), while with $\rdec,\ldec = 0$ we regain
the homogeneous transition. Notice that the barrier does not need to
be symmetric, i.e. $\ldec \neq \rdec$, with the extreme scenario being
a barrier with $\rjq = \matjump{\vps}{\vs}$ and $\ljq = 0$ yields a
one-way barrier or gate. In such a case the movement from $\vps$ to
$\vs$ is allowed but from $\vs$ to $\vps$ is not.

It is also possible to have dynamics opposite to the partially
reflecting barrier. In this case, again with $\vs$ and $\vps$ two
neighbouring sites, one has $\rjq = -\racc \matjump{\vs}{\vs}$ and
$\ljq=-\lacc \matjump{\vps}{\vps}$ where $\racc,\lacc \in [0, 1]$. As
the probability of jumping to the neighbours increases whilst the
probability of staying put decreases, we have chosen the name
anti-barrier for this type of heterogeneity.

\subsubsection{Long Range Connection}
\begin{figure}[H]
    \centering
    \includegraphics{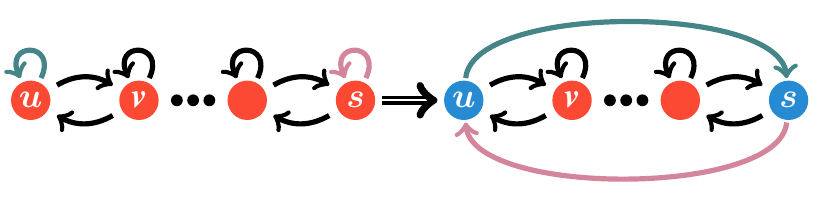}
    \caption{Example of a long range connection obtained by rewiring
      the lazy probability of $\vs$ and $\vps$ to create a long range
      connection between them (from left to right). In this case
      $\lacc=\lacc[][s]=1$.}
 \label{fig:example_defect_long}
\end{figure}

When adding an outgoing long-range connection one has to draw the
probability from one or more of the existing transitions. Let us
consider a site $\vs$ and a non-neighbouring destination site
$\mvec{s}$, where
$\matjump{\mvec{s}}{\vs} = \matjump{\vs}{\mvec{s}} = 0$. One way of
introducing the outgoing long range connection is to draw upon the
lazy (also called sojourn) probability using
$\jqsymbol_{\mvec{s}, \vs} =-\racc[][s] \matjump{\vs}{\vs}$ and
$\jqsymbol_{\vs,\mvec{s}} =-\lacc \matjump{\mvec{s}}{\mvec{s}}$, where
$\lacc, \racc[][s] \in [0, 1]$ is the proportion of the lazy
probability added to the long range connection, see
\cref{fig:example_defect_long} for a pictorial representation.

Note this is not the only way; one can also rewire an existing
connection from a neighbour to the non-neighbour. In such a case, with
$\vps$ a neighbour of $\vs$, we let $\rjq = \matjump{\vps}{\vs}$ and
$\rjq[][\mvec{s}] = -\matjump{\vps}{\vs}$. The former removes the
possibility of jumping from $\vs$ to the neighbour $\vps$, whilst the
latter adds the possibility of hopping from $\vs$ to the non-neighbour
$\mvec{s}$.
\subsubsection{Sticky and Slippery Sites}
\label{sec:sticky_slippery_params}
\begin{figure}[H]
  \centering
    \includegraphics{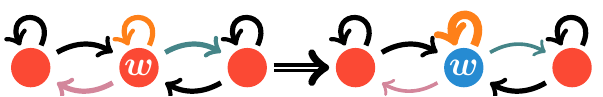}
    \caption{Example of a sticky site on a 1D lattice generated by
      reducing all of the outgoing probability to the neighbours as
      shown by the thinner arrows, whilst increasing the staying
      probability of $\ws$ as shown by the thicker self loop
      (from left to right).}
\label{fig:example_defect_sticky}
\end{figure}

Adding a partially reflecting barrier between two neighbouring sites
naturally increases the probability of staying. By harnessing this
property one can use multiple one-way partially reflecting barriers
between a site $\ws$ and all of its $k$ nearest neighbours,
$\mvec{r}_1, \cdots, \mvec{r}_k$, giving
$\rjq[][\mvec{r}_i][\ws]=\decsymbol_{}\matjump{\mvec{r}_i}{\ws}$,
and $\rjq[][\ws][\mvec{r}_i] = 0$ with $\decsymbol_{} \in [0, 1]$
for all $i = 1, \cdots, k$. The result is a sticky site, $\ws$,
where the probability of staying is increased, whilst the probability
of jumping to any of its neighbours is decreased. The introduction of
$\decsymbol$ is used to control and distribute the stickiness equally
across the neighbours in a convenient manner. See
\cref{fig:example_defect_sticky} for a pictorial representation on a
1D lattice.

Conversely, keeping $\rjq[][\ws][\mvec{r}_i] = 0$ and letting
$\rjq[][\mvec{r}_i][\ws]=-\frac{\accsymbol}{k}\matjump{\ws}{\ws}$ with
$\accsymbol \in [0, 1]$ for all $i = 1, \cdots, k$ yields a slippery
site with opposite dynamics. As for the sticky site, the introduction
of $\accsymbol$ is used to control the slippery quality of the site
$\ws$ equally among its neighbours. Note that we have chosen to divide
$\accsymbol$ by $k$ so that \cref{eq:lam_const_2} is automatically
satisfied.
\section{Heterogeneous Propagator}
\label{sec:heterogenous_propagator}

We consider an arbitrary collection of heterogeneous connections given
by a set of $M$ paired defective sites or defects,
$S = \lc \{\vs_1, \vps_1\}, \cdots,\{\vs_M, \vps_M\} \rc$. We use
$\vs_i$ and $\vps_i$ with subscripts to indicate the two members of
the $i^{\mathrm{th}}$ pair, while $\vs$ and $\vps$ without subscripts
refers to a generic pair in $S$. The pairs are unique, i.e.
$\lc \vs_i, \vps_i\rc \neq \lc \vs_j, \vps_j\rc$ for any $i \neq j$,
however, a site can be part of multiple pairs. For example, the set of
pairs which represents the schematic depicted in
\cref{fig:defect_schematic} is
$S = \lc \{\vs, \vps\}, \{\vs, \mvec{r}\}, \{\vps, \mvec{s}\}\rc$,
with the sites $\vs$ and $\vps$ being part of two pairs while the
sites $\mvec{r}$ and $\mvec{s}$ being part of only one pair each.  The
evolution of the occupation probability is given by the Master
equation (for the details see \cref {sup-sec:refl_defect_deriv} of
the Supplementary Materials)
\begin{equation}
  \begin{split}
&\rdpropt(\nvec, t + 1) = \sum_{\mvec{m}}\mmat{A}_{\nvec, \mvec{m}}\, \rdpropt(\mvec{m},  t) \\
&+ \sum_{k = 1}^{M}
\lb \delta_{\nvec, \vs_k} - \delta_{\nvec, \vps_k} \rb
\ls \bigg.  \rjq[k] \rdpropt(\vs_k, t) 
- \ljq[k]\rdpropt(\vps_k, t) \rs,
  \end{split}
\label{eq:rdprop_setup_full_master_main}
\end{equation}
where the second summation is over all pairs of heterogeneous
connections. When all $\jqsymbol$ parameters are set equal to zero,
\cref{eq:rdprop_setup_full_master_main} reduces to
\cref{eq:homogenous_master_eq} and the occupation probability on the
heterogeneous lattice, $\rdpropt{(\nvec, t)}$, reduces to that of the
homogeneous lattice, $\propt{(\nvec, t)}$.

One can find the generating function ($z$-domain) solution of
\cref{eq:rdprop_setup_full_master_main} by generalising the so-called
defect technique to obtain (see \cref{sup-sec:refl_defect_deriv} of
the Supplementary Materials)
\begin{equation}
\rdpropzz{\nvec}{\novec}  =
\propzz{\nvec}{\novec} - 1 + \frac{\mdet{\hcnmat[0]}}{\mdet{\hmat}},
  \label{eq:rdprop_sol_mt}
\end{equation}
where $\ztrans{f}(z) = \sum_{t = 0}^{\infty} f(t)z^t$ is the
generating function of the time dependent function $f(t)$,
$\propzz{\nvec}{\novec}$ is the propagator generating function of
\cref{eq:homogenous_master_eq}, while $\mdet{\hmat}$ and
$\mdet{\hcnmat[0]}$ are determinants with
\begin{align}
  &\hmat_{\matidx{i}{j}} = \rjq[i]  \propzzdif{\vs_i}{\vs_j}{\vps_j} \nonumber
  \label{eq:hmat_mt}
                          - \ljq[i] \propzzdif{\vps_i}{\vs_j}{\vps_j}  \\
                      &\phantom{\hmat_{\matidx{i}{j}}} - z^{-1}\delta_{i, j}, \\
  &\hcnmat[0]_{\matidx{i}{j}} = \hmat_{\matidx{i}{j}} \nonumber
  \label{eq:hcn0mat_mt}
                            - \propzzdif{\nvec}{\vs_j}{\vps_j}   \\
  &\phantom{\hcnmat[0]_{\matidx{i}{j}}}\times \ls \rjq[i]\propzz{\vs_i}{\novec} - \ljq[i]
    \propzz{\vps_i}{\novec} \rs.
\end{align}
In \cref{eq:hmat_mt,eq:hcn0mat_mt} we have used the notation
$\propzzdif{\nvec}{\vs}{\vps}=\propzzdifful{\nvec}{\vs}{\vps}$. From
here onwards we refer to $\propzz{\nvec}{\novec}$ as the homogeneous
propagator, which is known explicitly (see
\cref{sup-sec:block_matrix_const,sup-sec:eigvecs_eigvals_props} of
the Supplementary Materials), while $\rdpropzz{\nvec}{\novec}$ is referred
to as the heterogeneous propagator. When $t = 0$, that is $z = 0$, we
have $\propz_{\novec}{\lb \nvec, 0 \rb} = \delta_{\novec, \nvec}$,
while $\mdet{\hcnmat[0]} / \mdet{\hmat} = 1$ and we recover the
appropriate initial condition,
$\rdpropzsymbol_{\novec}{\lb \nvec, 0 \rb} = \delta_{\novec, \nvec}$.

In general the size of matricies $\hmat$ and $\hcmat[0]$ depends on
the number of paired defects, $M$. A $d$-dimensional walk with one
sticky (or slippery) site requires two paired defects for each of the
$d$ dimensions. However, in this case one can make a simplification
and reduce the size of the matrices by a factor of $2d$. The
simplified matrices $\hmat$ and $\hcmat[0]$ are defined, respectively,
in \cref{sup-eq:hmat_sticky,sup-eq:hcn0mat_sticky} of
the Supplementary Materials.

In \cref{fig:refl_defect_probability} we plot a snapshot of
$\rdpropz_{\novec}(\nvec, t)$ for the heterogeneities depicted in each
of the panels of \cref{fig:refl_schematic}.
\begin{figure}
  \centering
  \includegraphics{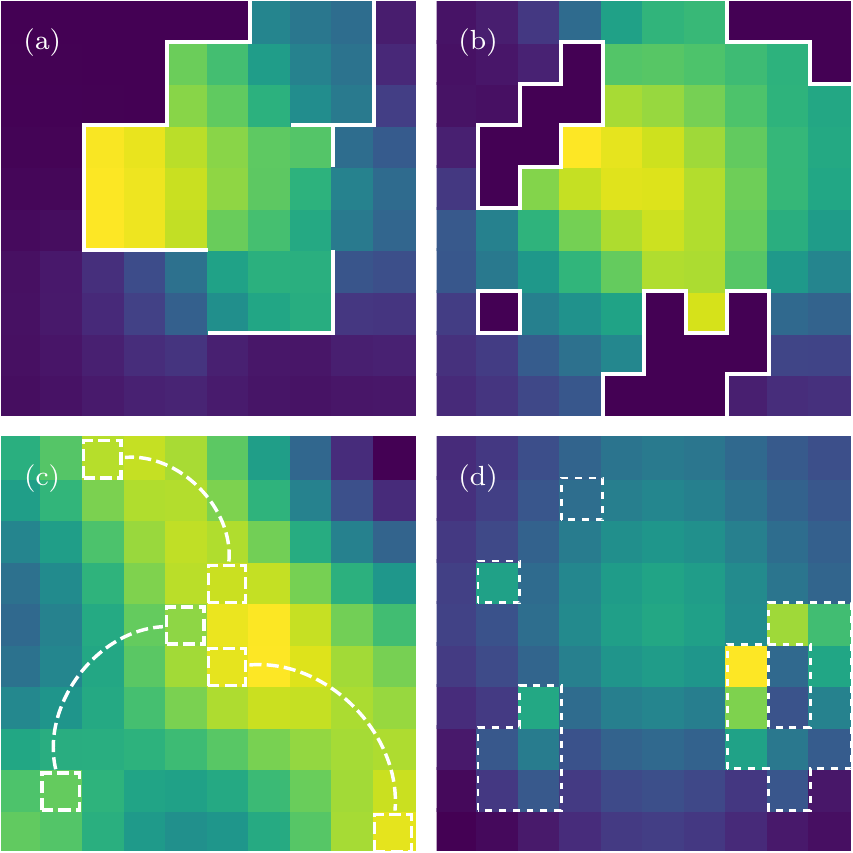}
  \caption{ A snapshot of $\rdproptt{\nvec}{\novec}$ at time $t = 100$
    obtained from \cref{eq:rdprop_sol_mt} with standard numerical
    methods \cite{abatewhitt1992, abateetal2000}.  Propagator with
    different configurations of defects, corresponding with
    \cref{fig:refl_schematic}, at $t = 100$. For all panels, the
    homogeneous propagator, $\propzz{\nvec}{\novec}$ is the 2D
    propagator with reflecting boundaries given in Eq.~(23) of
    Ref.~\cite{giuggioli2020} (see also
    \cref{sup-sec:eigvecs_eigvals_props} of
    the Supplementary Materials). The parameters used are: a
    domain of size $\mvec{N} = (10, 10)$, a localised initial
    condition with $\novec = (6, 6)$. We a use diffusion parameter of
    value $\mvec{q} =(0.2, 0.2)$, which gives the following transition
    probabilities: in the bulk of the homogeneous lattice the
    probability of jumping to one of the four neighbours is
    $\matjump{\mvec{r}}{\mvec{s}} = 0.05$ (with
    $\mvec{r} \neq \mvec{s}$), while the probability of staying at the
    same site is $\matjump{\mvec{r}}{\mvec{r}} = 0.8$. The reflecting
    barriers and other heterogeneities are super imposed on top of the
    probability. For panels (a) and (b), $\rjq = \matjump{\vps}{\vs}$
    and $\ljq = \matjump{\vps}{\vs}$ (for all
    $\lc \vs, \vps \rc \in S$) yielding perfectly reflecting
    barriers. For panel (c) $\rjq = -\frac{1}{2}\matjump{\vs}{\vs}$,
    and $\ljq = -\frac{1}{2}\matjump{\vps}{\vps}$. With this
    perturbation, when on one of the defective sites, the probability
    of staying is reduced to
    $\matjump{\vs}{\vs} = \matjump{\vps}{\vps} = 0.4$, while the
    probability of jumping to the non-neighbour is increased (from
    zero) to $\matjump{\vs}{\vps} = \matjump{\vps}{\vs} =
    0.4$. Lastly, for panel (d), for each of the sticky-sites $\ws$
    with $k$ neighbours $\mvec{r}_1,\cdots,\mvec{r}_k$ we use
    $\rjq[][\mvec{r}_i][\ws]= \frac{1}{4}\matjump{\mvec{r}_i}{\ws}$
    (see \cref{sec:sticky_slippery_params}). For convenience we have
    omitted colour bars for each panel as we are interested only in
    the relative differences of the occupation probability.}
    \label{fig:refl_defect_probability}
\end{figure}
In panel (a) the lattice is partitioned by impenetrable barriers
represented by the solid white lines. Here, one can observe the lowest
probabilities in the top-left corner since the walker has not had the
time to travel around the barriers. Panel (b) contains areas enclosed
by impenetrable barriers, with occupation probabilities that are
always zero. The long range connection shown in panel (c) has enabled
the walker to spread further than in other panels. Small peaks in the
probability can be observed away from the initial condition, in the
top-left, bottom-left and bottom right corners. In panel (d) the
sticky regions tend to show a higher occupation probability compared
to the homogeneous sites.
\begin{figure*}[ht]
  \centering
  \includegraphics{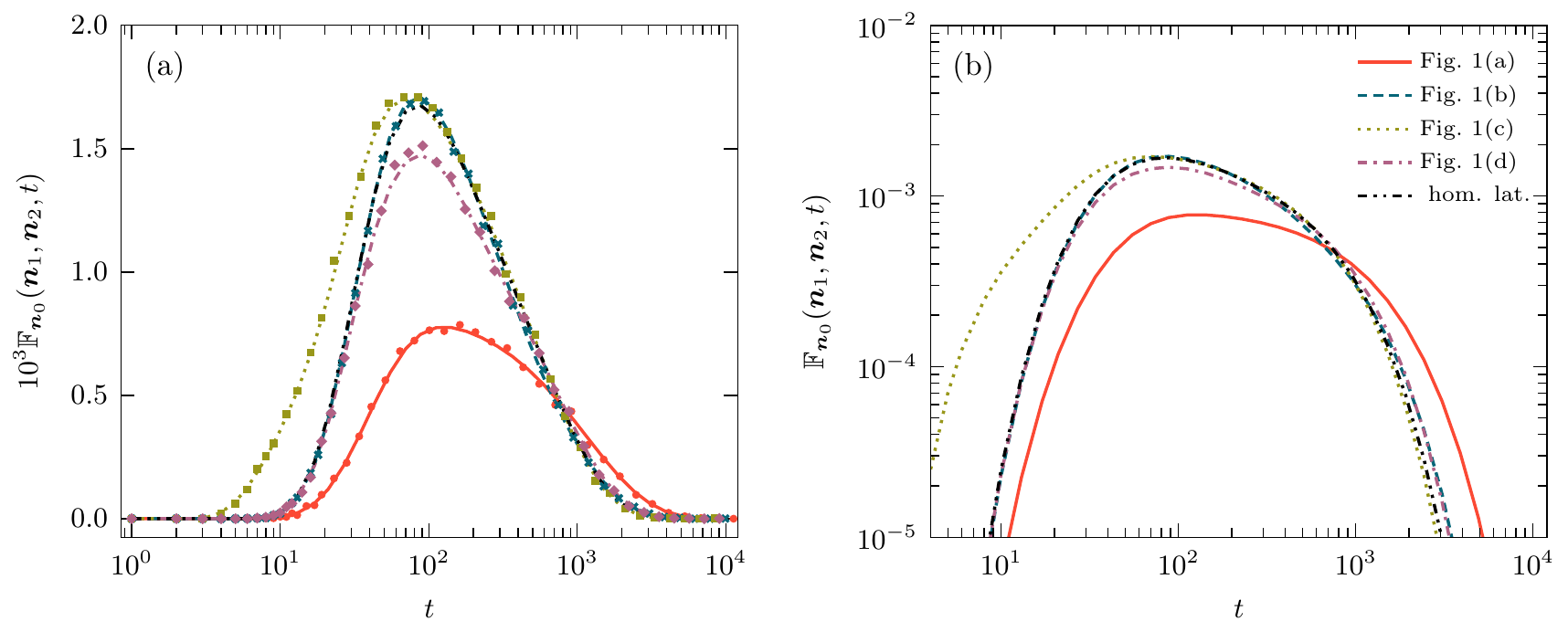}
  \caption{Time-dependent first-passage probability to either of two
    targets in the presence of heterogeneities. The location of the
    targets $\nvec_1 = (4, 2)$ and $\nvec_2 = (10, 7)$ in relation to
    the initial condition $\novec = (6, 6)$ and visualisation of the
    heterogeneities present can be seen in the schematic diagram in
    \cref{fig:refl_schematic}. We use a homogeneous propagator,
    $\propzz{\nvec}{\novec}$ with a reflecting domain of size
    $\mvec{N} = (10, 10)$ and a diffusion parameter of value
    $\mvec{q} = (0.2, 0.2)$. The explicit form of
    $\propzz{\nvec}{\novec}$ is given by Eq.~(23) of
    Ref.~\cite{giuggioli2020}. The lines are obtained through
    numerical inversion of the generating function of the
    first-passage probability to either of two targets (see text),
    while the corresponding marks---shown only in panel (a)---are
    obtained through $1.5\times10^{6}$ stochastic simulations.}
  \label{fig:first_passage_probability}
\end{figure*}

Note that we have not placed any restriction on whether (the
homogeneous propagator) $\propzz{\nvec}{\novec}$ conserves probability
or note. When there are fully or partially absorbing sites, one may
proceed in two ways. (i) In the first approach one account for the
absorbing dynamics by finding the propagator $\propzz{\nvec}{\novec}$
that satisfies appropriate boundary conditions, before adding inert
disorder via \cref{eq:rdprop_sol_mt}. (ii) In the second approach one
would take $\propzz{\nvec}{\novec}$ without any absorbing locations,
construct $\rdpropzz{\nvec}{\novec}$ and then add the absorbing sites
using the standard defect technique in the presence of absorbing sites
\cite{kenkrebook2021}. While the choice makes no impact on the final
dynamics, depending on the situation one procedure may be more
convenient than the other.

\section{First-Passage Processes}\label{sec:first_passage}
An important quantity derived from the propagators is the
first-passage statistics to a set of targets. It is relevant to
stochastic search in movement ecology \cite{nathangiuggioli2013},
swarm robotics \cite{giuggiolietal2018} and many other areas
\cite{redneretalbook2014}.

The first-passage probability, $\rdfptt{\nvec}{\novec}$, that is the
probability to reach $\nvec$ for the first time at $t$ having started
at $\novec$, is related to the propagator, $\rdproptt{\nvec}{\novec}$
by the renewal equation.  When $\nvec \neq \novec$, the well-known
relation in $z$-domain is given by
$\rdfpzz{\nvec}{\novec} = \rdpropzz{\nvec}{\novec}
/\rdpropzz{\nvec}{\nvec}$. Having the first-passage probability in
closed form allows one to substitute the heterogeneous first-passage
probability $\rdfptt{n}{n_0}$ (or $\rdfpzz{n}{n_0}$) in place of the
homogeneous counterpart in other established contexts where
homogeneous space was previously assumed.

One such context is a first-passage in the presence of multiple
targets, where one is interested in the probability of being absorbed
at any of the targets. We use recent findings \cite{giuggioli2020} to
determine the dynamics of a lattice walker to reach either of two
sites, $\nvec_1$, and $\nvec_2$, for the first time at $t$ in the
presence of spatial heterogeneities, given by
$\rdfptt{\nvec_1, \nvec_2}{\novec}$.  The generating function of this
probability, given by
$\rdfpzz{\nvec_1, \nvec_2}{\novec} = \lc \rdfpzz{\nvec_1}{\novec} \ls
- \rdfpzz{\nvec_2}{\nvec_1} \rs + \rdfpzz{\nvec_2}{\novec} \ls -
\rdfpzz{\nvec_1}{\nvec_2} \rs \rc \\ \times \ls 1 -
\rdfpzz{\nvec_2}{\nvec_1} \rdfpzz{\nvec_2}{\nvec_1} \rs^{-1}$, (taken
from Eq.~(38) in Ref.~\cite{giuggioli2020}), is expressed in terms of
the first-passage probabilities to single targets.

In \cref{fig:first_passage_probability} plot the time dependent
probability for the heterogeneity examples shown in
\cref{fig:refl_schematic}. The first non-zero probability corresponds
to the length of the shortest path to either of the targets, which in
the absence of heterogeneities and for the examples in
\cref{fig:refl_schematic}(a), (b) and (d) is 6, whereas for the panel
(c) one can reach the target $\nvec_1$ from $\novec$ in 4 steps as a
result of the nearest long range connection. It is clearly visible in
the earlier rise of the curve related to
\cref{fig:refl_schematic}(c). Interestingly, the first-passage
probability curves corresponding with excluded regions, shown in
\cref{fig:refl_schematic}(b), and the homogeneous case are almost
indistinguishable from each other for 2 decades. While excluding parts
of the lattice increases the lengths of some paths to the targets, it
also reduces the overall space that can be explored. For the setup
chosen, these two effects counteract each other at short and
intermediate timescales.

Among all the curves, the case with open partitions related to
\cref{fig:refl_schematic}(a), results in a first-passage probability
which is the slowest to rise and with the broadest tail in the
distribution. The reasons for such characteristics compared to all
other curves is due to the location of the initial site relative to
the targets. As the latter ones are partially behind partitions, the
more directed paths take more time to reach the targets and the walker
remains confined in the region around the initial site for much
longer.

The sticky sites in \cref{fig:refl_schematic}(d) have limited effect
on the more directed paths connecting the starting site and the
targets. This is why $\rdfptt{\nvec_1, \nvec_2}{\novec}$ in
\cref{fig:first_passage_probability}(b) is identical to the
homogeneous case at short times. However, sticky-sites can be both a
hindrance or a benefit to the searcher. While it can partially trap
the walker and stop it from reaching the target site, it can also stop
the walker from exploring regions away from the targets. Since there
are sticky sites close to the targets, these two effects counteract
one another and we observe marginal difference in the tail of the
distribution when compared with the homogeneous curve.

\subsection{Explicit mean first-passage quantities}
The first moment of $\rdfptt{\nvec}{\novec}$, that is the mean
first-passage time (MFPT),
$ \rdmfptt{\nvec}{\novec} = \ld \frac{\mathrm{d}}{\mathrm{d} z} \rdfpzz{\nvec}{\novec} \rp_{z = 1}$,
is given by
\begin{equation}
\rdmfptt{\nvec}{\novec}
= \frac {\mfptt{\nvec}{\novec} \mdet{\hmatd - 1/\mfptt{\nvec}{\novec}\hsmatd[1]}}
{\mdet{\hmatd - \hsmatd[2]}},
  \label{eq:rdmfpt_mt}
\end{equation}
where $\mfptt{\nvec}{\novec}$ is the homogeneous MFPT from $\novec$ to
$\nvec$ and the elements of the matrices $\hmatd$, $\hsmatd[1]$ and
$\hsmatd[2]$ are defined in terms of homogeneous MFPTs. They are
given, respectively, in
\cref{eq:mean_hij_final,eq:mean_h1ij_final,eq:mean_h2ij_final} for
general heterogeneities, while for the case with only sticky or
slippery heterogeneities the matrices are given by
\cref{sup-eq:mean_hij_1_sticky,sup-eq:mean_h1ij_1_sticky,sup-eq:mean_h2ij_1_sticky}
in the Supplementary Materials. In the coming sections we
use the mathfrak notation e.g. $\rdmfptsymbol,\rdmretsymbol$ and
$\rdmexitsymbol$, for statistics involving the heterogeneous dynamics,
while the mathcal notation, e.g.  $\mfptsymbol, \mretsymbol$ and
$\mexitsymbol$, is reserved for the homogeneous counterpart. The
dependence on the target at $\nvec$ is only present in the matrices
$\hsmatd[1]$ and $\hsmatd[2]$; the dependence on the initial condition
$\novec$ is only present in $\hsmatd[1]$; and the dependence on the
location of the heterogeneities are in all three matrices.

The probability distribution of the first-return time is related to
the propagator
via $\rdretzz{\nvec} = 1 - \rdpropzz[-1]{\nvec}{\nvec}$, with a mean
return time (MRT) given by
\begin{equation}
  \label{eq:rdmret_mt}
\rdmrett{\nvec}
=
\frac
{
\mrett{\nvec}
\mdet{\hmatd}
}
{
\mdet{\hmatd - \hsmatd[2]}
},
\end{equation}
where $\mrett{\nvec}$ is the homogeneous mean return time.

When the heterogeneities preserve the symmetric properties of the
homogeneous lattice, i.e. the disorder does not add any bias to a
diffusive system or remove any bias present in a system with drift,
then the ratio
$ \small {\matjump{\vs}{\vps}}/{\matjump{\vps}{\vs}} = {\lb
  \matjump{\vs}{\vps} - \ljq \rb}/{\lb \matjump{\vps}{\vs} - \rjq \rb}
$ is satisfied, for all $\{\vs, \vps\} \in S$, and the heterogeneous
system maintains the steady state of the homogeneous system. In this
case, $\hsmatd[2] = 0$, the MFPT given by \cref{eq:rdmfpt_mt}, can be
simplified to
$\rdmfptt{\nvec}{\novec} = \mfptt{\nvec}{\novec} - 1 + \mdet{\hmatd -
  \hsmatd[1]} / \mdet{\hmatd}$, while the MRT remains the same as the
homogeneous MRT, $\rdmrett{\nvec} = \mrett{\nvec}$ as expected from
Kac's lemma \cite{kac1947a}. (see \cref{ap:mret} and
\cref{sup-sec:mret_deriv} of the Supplementary Materials).

In the presence of multiple targets at the outer boundary of the
domain, we relate the first-passage probability to any of the targets
to a propagator with the appropriate absorbing boundaries. In this
case, the first-passage is referred to as the first-exit, and its
probability generating function is related to the propagator through
the relation $\rdexitzz{\novec} = 1 - (1 - z) \rdsurvzz{\novec}$,
where $\rdsurvzz{\novec}$ is the survival probability. Taking the mean
of the distribution (see \cref{ap:mexit_deriv}) gives
\begin{equation}
  \label{eq:rdmexit_mt}
  \rdmexitt{\novec} = \rdsurvzsymbol_{\novec}(z = 1) =
  \ld \frac{\mexitt{\novec} \mdet{\hmat - 1/\mexitt{\novec}\hsurvmat}}{\mdet{\hmat}} \rp_{z = 1}
\end{equation}
where $\mexitt{\novec}$ is the mean exit time starting at $\novec$
without any heterogeneities, and the matrix $\hsurvmat$ is given
explicitly in \cref{eq:hsurv_mat}.  The presence of one or more
absorbing boundaries on the homogeneous propagator
$\propzz{\mvec{s}}{\mvec{r}}$ allows for a simple evaluation at
$z = 1$. That is to say $\propzsymbol_{\mvec{r}}{(\mvec{s}, z = 1)}$
is finite for any $\mvec{r}$ and $\mvec{s}$ in the domain; and
therefore $\hmat$ and $\hsurvmat$ also remains finite and can be
easily evaluated.

\begin{figure}[ht]
  \centering
  \includegraphics{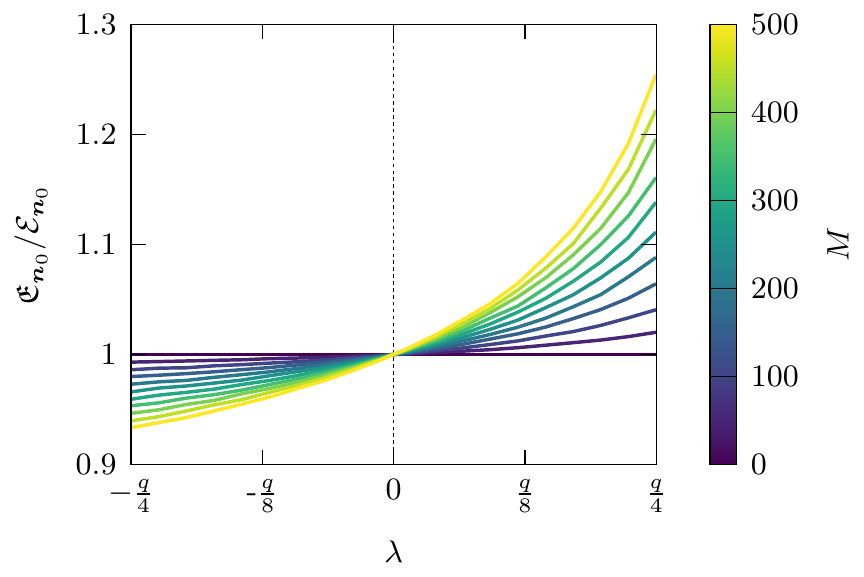}
  \caption{\label{fig:mexit}The ratio of the heterogeneous mean exit
    time $\rdmexitt{\novec}$ to the homogeneous mean exit time
    $\mexitt{\novec}$ for randomly distributed barriers and
    anti-barriers. We use a homogeneous propagator with a domain of
    size $\mvec{N} = (51, 51)$ with absorbing boundary conditions, an
    initial condition at the centre of the domain $\novec = (26, 26)$
    and a diffusion parameter of $\mvec{q} = (0.8, 0.8)$. The explicit
    form of $\propzz{\nvec}{\novec}$ is given by the $z$-transform of
    Eq.~(23) of Ref.~\cite{giuggioli2020}. Each curve is obtained
    using \cref{eq:rdmexit_mt} and performing an ensemble average with
    $10^2$ sample realisations of locations of barriers
    ($\jqsymbol > 0$) or anti-barriers ($\jqsymbol > 0$) for each
    $\jqsymbol$.}
\end{figure}

In \cref{fig:mexit} we show the effect of randomly distributed
barriers and anti-barriers as a function of the barrier strength in a
2D domain with absorbing boundaries. The $M$ neighbouring defective
site pairs are uniformly distributed on the lattice with
$\jqsymbol = \rjq = \ljq$ for all $\lc \vs, \vps \rc \in S$.
One can see that for $\jqsymbol > 0$, $\rdmexitt{\novec}$ increases,
as the heterogeneous connections behave as a partially reflecting
barrier slowing down the walker. Furthermore an increase in the number
of heterogeneities resulting in larger exit times. Conversely, when
$\jqsymbol < 0$ the heterogeneous connections become anti-barriers
increasing the probability of jumping across compared to the
homogeneous case, which effectively increases the spread of the walker
leading to shorter exit times. When the barriers are impenetrable,
increasing the number of barriers also increases the likelihood of the
walker being trapped and unable to reach the boundary and will cause
the MET to diverge. Although we do not study it here, a similar setup
could be used to analyse percolation in finite multidimensional
domains.

\subsection{First-passage processes in 1D with a single barrier and
  the phenomenon of disorder indifference of the
  MFPT}\label{sec:theoretical_analysis}

We consider a simple spatial heterogeneity in a 1D domain with a
partially reflecting barrier between $u$ and $u+1$. To study the
dependence of the position and strength of the barrier (or
anti-barrier) on the first-passage dynamics, we first fix the position
of target and initial sites with $n> n_0$; assume a reflecting
boundary between $n=0$ and $n=1$; and take
$\jqsymbol_{u, u+1} = \jqsymbol_{u+1, u} = \jqsymbol$ with
$\jqsymbol \in [-(1 - q), q/2]$. In this case the first passage
probability can be written using the convenient notation
\begin{equation}
  \label{eq:fpz_1d_1_bar}
  \begin{aligned}
    \rdfpzz{n}{n_0} =
    \lc
    \begin{array}{ll}
      \frac{a{\lb n_0, z \rb} - \frac{2\jqsymbol}{q}b{\lb n_0, u, z \rb}}{a{\lb n, z\rb} - \frac{2\jqsymbol}{q}b{\lb n, u, z \rb}} & u < n_0 \\
      & \\
      \frac{a{\lb n_0, z \rb} - \frac{2\jqsymbol}{q}a{\lb n_0, z \rb}}{a{\lb n, z\rb} - \frac{2\jqsymbol}{q}b{\lb n , u, z\rb}} & u \geq n_0 
      \end{array} \rd,
  \end{aligned}
\end{equation}
where
$a(n, z) = \cosh{\ls \lb \frac{1}{2} - n \rb \zeta \rs} \cosh{\ls
  \frac{1}{2} \zeta \rs}$,
$b(n, u, z) = \cosh{\ls \lb 1 - n \rb \zeta \rs} + \sinh{\ls \lb n -
  2u - \frac{1}{2} \rb \zeta \rs}\sinh{\ls \frac{1}{2} \zeta \rs}$,
$\zeta = \acosh{\ls 1 - \frac{1}{q} \lb 1- \frac{1}{z} \rb \rs}$, and
with the probability of moving given by $q \in (0, 1]$. The
homogeneous first-passage probability, \fpzz{n}{n_0}, can be recovered
from \cref{eq:fpz_1d_1_bar} by letting $\jqsymbol \to 0$. When the
barrier is to the left of the initial condition, the limit
$\jqsymbol \to \frac{q}{2}$ creates an impenetrable barrier, the
behaviour is equivalent to shifting both the target and the initial
condition to the left by $u$ giving
$\rdfpzz{n}{n_0} = \fpzz{n - u}{n_0 - u}$. Whereas, when
$n_0 \leq u < n$, the same limit gives $\rdfpzz{n}{n_0} = 0$ as the
walker becomes blocked by the barrier and can never reach the target.
\begin{figure}[ht]
  \centering
  \includegraphics{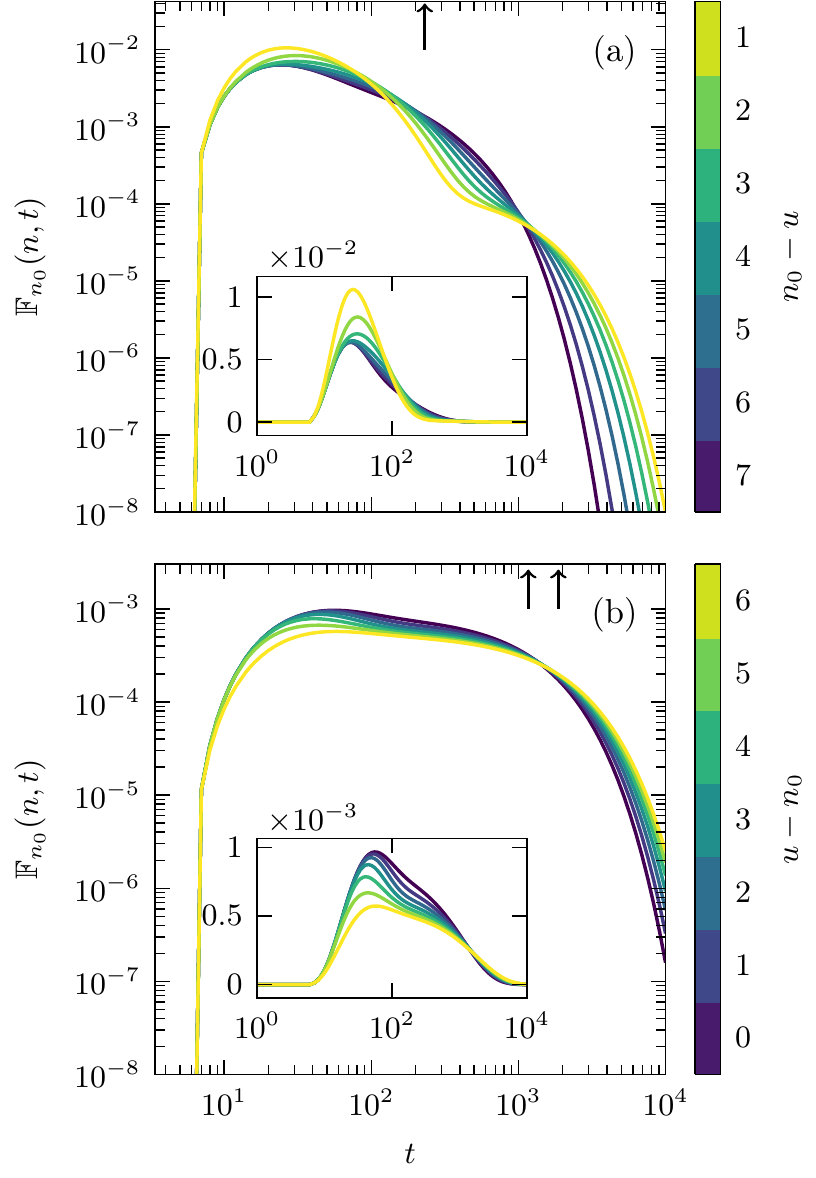}
  \caption{\label{fig:first_passage_1d}The time dependent first
    passage probability given through the numerical inversion of
    \cref{eq:fpz_1d_1_bar}. Panel (a) represents the scenario
    $u < n_0 < n$, while panel (b) is the case when $n_0 \leq u <
    n$. The values of the other parameters are:
    $\jqsymbol = 0.975 \cdot \frac{q}{2}$; a diffusion parameter of
    value $q = \frac{2}{3}$; the initial condition $n_0 = 8$; the
    target site $n=15$; and a reflecting boundary between $n=0$ and
    $n=1$. The arrows indicate the MFPTs: in panel (a) all of the
    curves have the same MFPT of $\rdmfptt{n}{n_0} = 231$ (disorder
    indifference), whereas in panel (b) the two arrows indicate the
    minimum and maximum MFPTs of the curves which are
    $\rdmfptt{n}{n_0} = 1167$ and $\rdmfptt{n}{n_0} = 1869$, and
    obtained, respectively, when $u=8$ and $u=14$ }
\end{figure}

In \cref{fig:first_passage_1d}, we plot the time dependence of
\cref{eq:fpz_1d_1_bar} for the two different scenarios $u < n_0$ and
$u \geq n_0$ represented, respectively, by panels (a) and (b). With
$u < n_0$ and $\jqsymbol < q/2$, that is the barrier to the left of
the initial condition, as one increases $u$ from $u=1$ one observes an
increase in the modal peak. When the walker is reflected by the
permeable barrier, it stops the walker from straying further left and
effectively reduces the space that can be explored, increasing the
probability of reaching the target at an earlier time. However, if the
walker passes through the barrier, the partial reflection dynamics
becomes a hindrance: the walker is kept in the range $[1, u]$, causing
the probability of reaching the target at long times to increase
also. As probability in the tail and the mode increases, the
probability conserving $\rdfptt{n}{n_0}$ demands a reduction at
intermediate times, which is clearly visible from the figure. This
permeability induced mode-tail enhancement can also be witnessed by
fixing $u$ and changing $\jqsymbol \in [0, q/2)$, and we have also
observed the inverse effect, mode-tail compression by having
anti-barrier with $\jqsymbol \in [q-1, 0]$. We have chosen not display
these latter cases for want of space. Similar features have been
observed in a diffusing diffusivity model in
Ref.~\cite{lanoiseleeetal2018}, where increases in the probability at
short and long timescales were attributed to the dynamic diffusivity.
Our findings point to the fact that such richness can also emerge from
a static disorder at a single location.

Differently from the case when the barrier is to the left of the
initial condition, is the case when $u \geq n_0$ . In this scenario,
the barrier is always acting to slow the search process down, reducing
the probability of reaching the target at early times and increasing
the probability at long times as seen by the flattening of the mode
and the broadening of the tail, as shown in
\cref{fig:first_passage_1d}(b).

Computing the mean via either \cref{eq:fpz_1d_1_bar} or from
simplifying \cref{eq:rdmfpt_mt} yields the compact expression
\begin{equation}
  \label{eq:rdmfpt_ref_diff_1d}
  \rdmfptt{n}{n_0} = \mfptt{n}{n_0} + \frac{2}{q} \frac{\jqsymbol}{\frac{q}{2} -\jqsymbol}\lc 
  \begin{array}{ll}
    0, & u < n_0,\\
    u, & u \geq  n_0,
  \end{array}
  \rd
\end{equation}
where $\mfptt{n}{n_0} = (n - n_0) (n + n_0 - 1) / q$ is the 1D
homogeneous MFPT for $n_0 \leq n$ (given by Eq.~(14) of
Ref.~\cite{giuggioli2020}). Astonishingly, the mode-tail enhancement
present in the time-dependent probability when $u < n_0$ has no effect
on the mean. This is what we have termed the disorder indifference
phenomenon.

To explain why there is such an effect of disorder indifference, we
split the first passage trajectories into two mutually exclusive
subsets: the trajectories that never return to the initial condition
before reaching the target site on the right and those that return at
least once before reaching the target site. Clearly, the former
trajectories are unaffected by the presence of a barrier. The latter
trajectories \textit{can} be affected by the barrier, however, in
computing the mean one deals with mean return times which are
unaffected from the homogeneous case when
$\rjq[][u][u+1] = \ljq[][u+1][u]$ as stated in the previous section
(see \cref{sec:one_barrier} for the mathematical details).

An analogue of this indifference phenomenon was observed in
Ref.~\cite{godecmetzler2015}, where the MFPT in a quasi-1D domain in
continuous space with two layers of different diffusivity was
studied. When the initial condition was in between the interface of
the layers and the target, they observed that the MFPT was indifferent
to the diffusivity of the media beyond the interface. In that study,
the first-passage probability was not considered and the cause of this
indifference could not be quantified. However, one can relate the
location of the interface of their system with the position of the
barrier in ours. Through this relation, we believe that the behaviour
observed in Ref.~\cite{godecmetzler2015}, is closely related to the
dynamics presented in \cref{fig:first_passage_1d}.

Given a barrier between the initial condition $n_0$ and the target
$n$, the effect on the MFPT increases linearly as the displacement
from the boundary increases. While the effect, which can be to speed
up ($\jqsymbol < 0$) or to slow down ($\jqsymbol > 0$), is due to the
disorder, the linear dependence is not. This linear dependence is
present in all 1D situations and is proportional to the distance
between the initial condition and the reflecting boundary (see
\cref{sec:linear_dep_disorder}).

To explore the effects of asymmetry in the heterogeneities we consider
the MRT with $\rjq[][u+1][u]\neq\ljq[][u][u+1]$. In this cases, the
steady state is no longer homogeneous, effectively creating an
out-of-equilibrium system due to the loss of detailed balance in the
Master equation. To illustrate this point, we consider the MRT of a 1D
walker within a segment of length $N$ with reflecting boundaries and
with a barrier between $u$ and $u+1$,
($\rjq[][u+1][u] \neq \ljq[][u][u+1]$). In this case,
\cref{eq:rdmret_mt} simplifies to
\begin{equation}
  \label{eq:mret_1d_ref}
\rdmrett{n} = 
\lc
\begin{array}{ll}
N\ls \frac{\ffrac{q}{2}-\rjq[][u+1][u]}{\ffrac{q}{2} - \ljq[][u][u+1] } \rs
- u \ls \frac{\ljq[][u][u+1] -\rjq[][u+1][u]}{\ffrac{q}{2}- \ljq[][u][u+1]} \rs, & n <  u + 1,  \\
& \\
N - u \ls \frac{\rjq[][u][u+1] - \ljq[][u+1][u]}{\ffrac{q}{2}-\rjq[][u+1][u] } \rs, & n \geq u + 1, 
\end{array} 
\rd.
\end{equation}
One can see that when $\rjq[][u+1][u] = \ljq[][u][u+1]$, the MRT
reduces to $N$ regardless of whether $n \leq u$ or $n > u$. In the
extreme case, where the barrier is impenetrable in both direction,
$\rjq[][u+1][u] = \ljq[][u][u+1] = q/2$, one can recover the
appropriate MRTs when $n \leq u$ and $n > u$, which are, respectively,
$u$ and $N - u$ (see \cref{sup-sec:mret_limit} of the Supplementary
Materials).

Thus far we have focused on technical development and theoretical
insights. As we move forward, the remainder of the article is devoted
to practical examples, and is used to demonstrate the applicability of
the framework.  For practical convenience the details of the modelling
set up are given in the appendices and only the results are discussed.

\section{Transdermal Drug Delivery}
\label{sec:transdermal}

In the first application we consider the problem of optimising
transdermal drug delivery, that is the transfer of drugs through the
skin. One of the challenges of transdermal drug delivery is traversal
of the outer-most layer of the epidermis called the stratum corneum
(SC) by hydrophilic molecules \cite{elias2007}. This layer is made up
of dead cells called corneocytes which are arranged in a dense
`brick-and-mortar' like pattern \cite{nemessteinert1999}. Inspired by
some of the recent strategies proposed to enhance drug absorption
\cite{ramadonetal2021}, we consider the use of a micro-needles to
pierce first the SC before applying a drug patch. We study the
effectiveness of this method by using our modelling framework to
represent the SC as heterogeneities on a lattice and modelling the
movement of drug molecules as a random walk.
\begin{figure}[ht]
  \centering
  \includegraphics[width=\columnwidth]{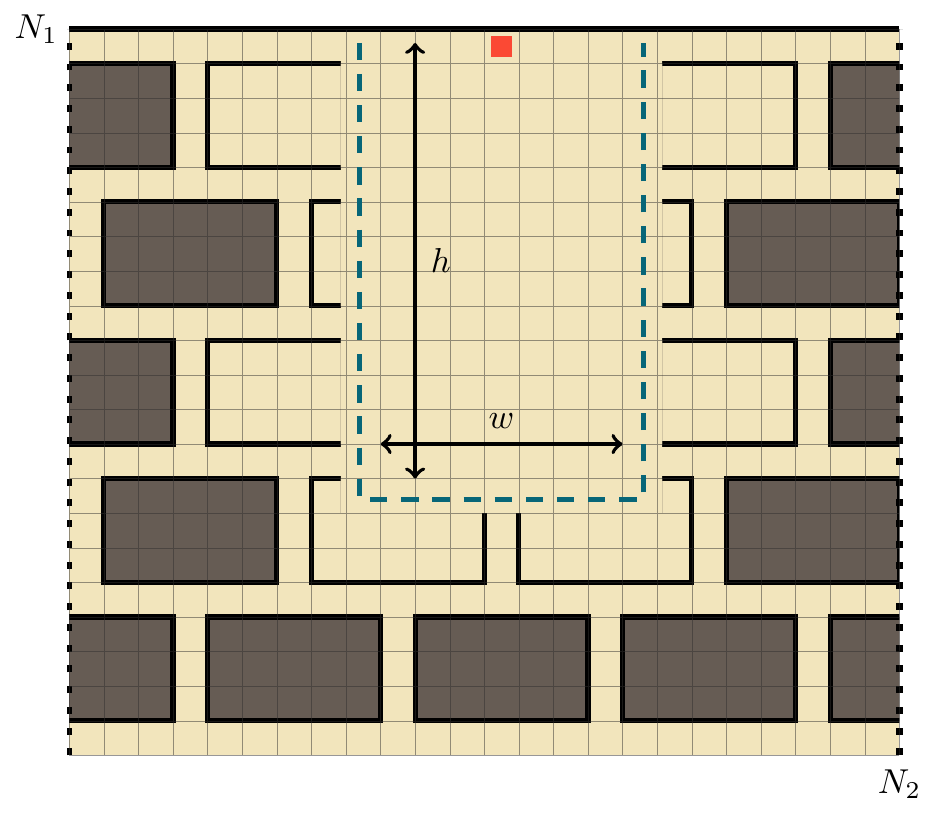}
  \caption{ Representation of the `brick-and-mortar' arrangement of
    the corneocytes in the stratum corneum (SC). The red square depicts
    the starting location of the random walker. The geometry is given
    by an an absorbing boundary at $n_1 = 1$, a reflecting boundary at
    $n_2 = N_1$, shown as a thick solid black line, and a periodic
    boundary in the second dimension, depicted as dashed black
    lines. By using a number of paired defects, one is able to cordon
    off sites (shaded grey), creating the `brick-and-mortar' pattern
    of the SC. The dashed blue rectangle with width $w$ and height $h$
    models the destruction of the SC structure via a micro-needle
    puncture, with $h$ and $w$ representing, respectively, the
    puncture height and width. This destruction may open up some of
    the `bricks', allowing the walker to easily travel inside. The
    initial position of the walker is at the centre-top of the
    puncture, $\novec = (N_1, N_2 / 2)$.}
  \label{fig:brick_and_mortar}
\end{figure}

We use a homogeneous 2D nearest-neighbour random walker subject to
mixed boundary conditions: an absorbing boundary located at $n_1 = 1$
and a reflecting boundary located at $n_1 = N_1$, for the first
dimension and a periodic boundary condition on the second
dimension. The heterogeneities are impenetrable barriers representing
the lipid matrix. These are arranged in a manner to create excluded
regions that form the `brick-and-mortar pattern of the SC, see
\cref{fig:brick_and_mortar}. The pattern is partially destroyed to
represent the needle piercing in a rectangle with height $h$ and width
$w$ resulting in an area absent of barriers as shown by the blue
dashed rectangle in \cref{fig:brick_and_mortar}.

The quantity of interest is the MET with an initial condition starting
at the reflecting end of the domain.
\begin{figure}[ht]
  \centering
  \includegraphics{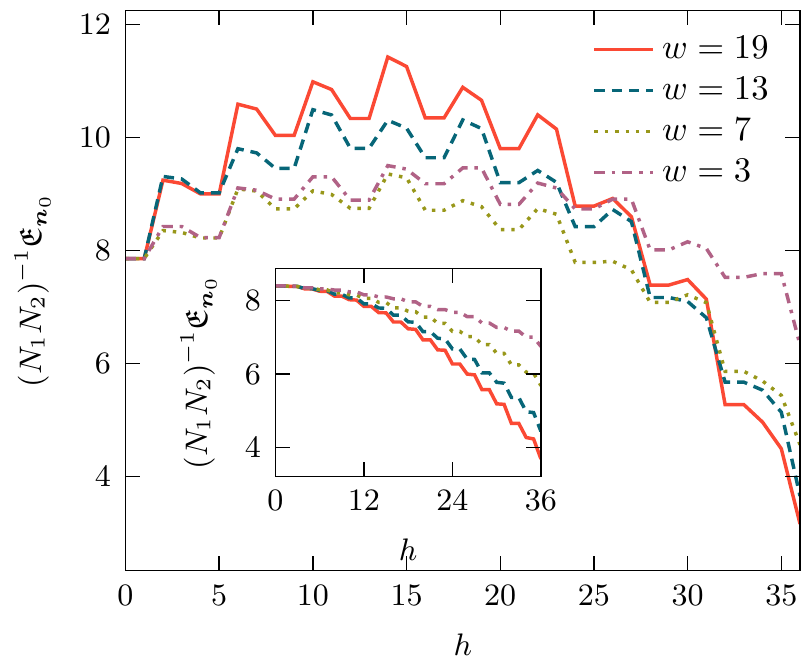}
  \caption{Mean exit time as a function of the puncture height, $h$,
    for different values of puncture width $w$ (see
    \cref{fig:brick_and_mortar} for description of the setup). We use
    a rectangular domain of size $\mvec{N} = (37,36)$, a diffusion
    parameter of $\mvec{q} = (0.8, 0.8)$, `bricks' of size $(3, 5)$
    resulting in 9 layers with 6 bricks per layer, an initial
    condition of $\novec = (1, 19)$. The main panel depicts the
    scenario where the barriers encapsulating the `bricks' are
    impenetrable, i.e. $\rdec = \rdec = 1$ leading to
    $\rjq = \ljq = 0.2$ for all $\vs, \vps$, while the inset shows the
    scenario where the barriers are partially permeable with
    $\rdec = \rdec = 1$ giving $\rjq = \ljq = 0.18$.}
  \label{fig:transdermal_bore}
\end{figure}
We plot the MET as a function of the puncture depth and width in
\cref{fig:transdermal_bore}. The overarching qualitative changes in
the MET can be explained by two competing effects. The first is the
breaking of enclosed bricks to create open partitions. The additional
sites available for exploration makes the paths to reach the absorbing
boundary longer.

The second effect is that the puncture allows the walker more direct
movement towards the bottom layers leading to smaller MET. The removal
of some of the impenetrable barriers allows for more direct paths to
the absorbing boundary, which leads to smaller mean exit times. The
strength of this effect is dependent on the size of the puncture
$hw$. For small values of $h$, the first effect has greater influence
leading to an increase in the METs. As $h$ is increased the second
effect becomes more prominent and drives down the METs resulting in a
global maximum. The interplay between the two effects also gives rise
to the oscillations. Puncture of a brick layer opens it up, leading to
larger exit times as the walker becomes temporarily confined inside a
brick. Increasing the puncture height further destroys the brick
structure of a layer and allows the walker to traverse the latter via
a direct route thereby decreasing the exit times.

The global maximum and the oscillations are only present when the
barriers are highly reflecting or impenetrable i.e.
$0 \ll \rjq \ljq \leq q/2$ for all $\lc \vs, \vps\rc \in S$.  The
maximum is lost when the permeability gets larger as the random walker
is only partially confined by the barriers, leading to a monotonic
decrease in the MET as seen in the inset of
\cref{fig:transdermal_bore}. With permeable barriers all the sites are
always accessible independently of $h$ and $w$, puncturing only creates
more direct routes to the absorbing boundary leading to smaller exit
times.

\section{Thigmotaxis}
\label{sec:thigmotaxis}
For the second application we look at thigmotaxis, which broadly
speaking, is the movement of an organism due to a touch stimulus. We
are interested specifically in the tendency of animals to remain close
to the walls of an environment, a behaviour that is observed in many
species from insects to mammals \cite{higakietal2018,doriaetal2019}.
\begin{figure}[ht]
  \centering
  \includegraphics{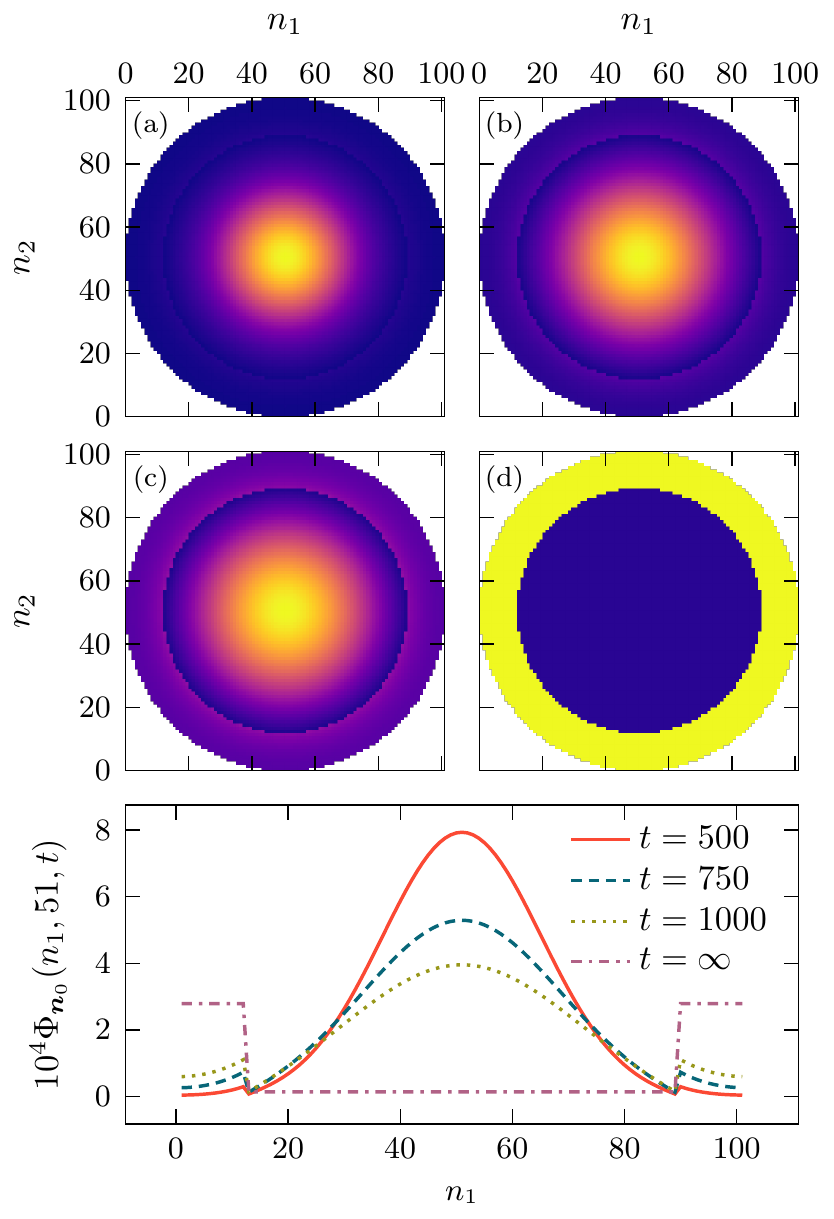}
  \caption{The propagator $\rdproptt{\nvec}{\novec}$ at different
    moments in time, $t=500, 750, 1000, \infty$ where the walker is
    initially at the centre of the domain, $\novec = (51, 51)$. When
    inside the inner region the walker can freely enter the outer
    region without any resistance, that is $\rjq = 0$ and when in the
    outer region the probability to move inward is modified via
    $\ljq = \decsymbol_{{i}}\matjump{\vps}{\vs}$. Other parameters
    used are the diffusivity of value $\mvec{q} = (0.8, 0.8)$ and a
    square domain of size $\mvec{N} = (101, 101)$, (see
    \cref{sec:thigmotaxis_defects} for details on the placement of
    defects)}
  \label{fig:thigomotaxis_probability} 
\end{figure}
We quantify the thigomotactic tendency by appropriately selecting
defects location and $\jqsymbol$ to represent regions which are more
easily accessible when moving in one direction (approaching
boundaries) versus another (moving away from boundaries).

Since we are able to construct arbitrary shapes with the formalism, we
consider two concentric circles within a square domain.  The first is
used to restrict the walker to a circular reflecting domain of radius
$R$. The second has a radius $r$, with $r$ < $R$, and is used to
partition the domain into two regions: an inner region; and an outer
region, which is the annulus between $r$ and $R$, representing the
preferred area of occupation. By placing one-way partially reflecting
barriers along the radius $r$, we allow the walker to leave the inner
region to enter the outer region without any resistance, while the
tendency of remaining in the outer region is controlled by the
parameter $\decsymbol_{i} \in [0, 1]$. With $\decsymbol_{i} = 1$ the
walker never leaves the outer region once it gets there, whereas with
$\decsymbol_{i} = 0$, the partially reflecting barrier are removed and
all areas of the circular domain becomes easily accessible. For a
details on the placement of the defects and the construction of the
circular domain see \cref{sec:thigmotaxis_defects}.
\begin{figure}[ht]
  \centering
  \includegraphics{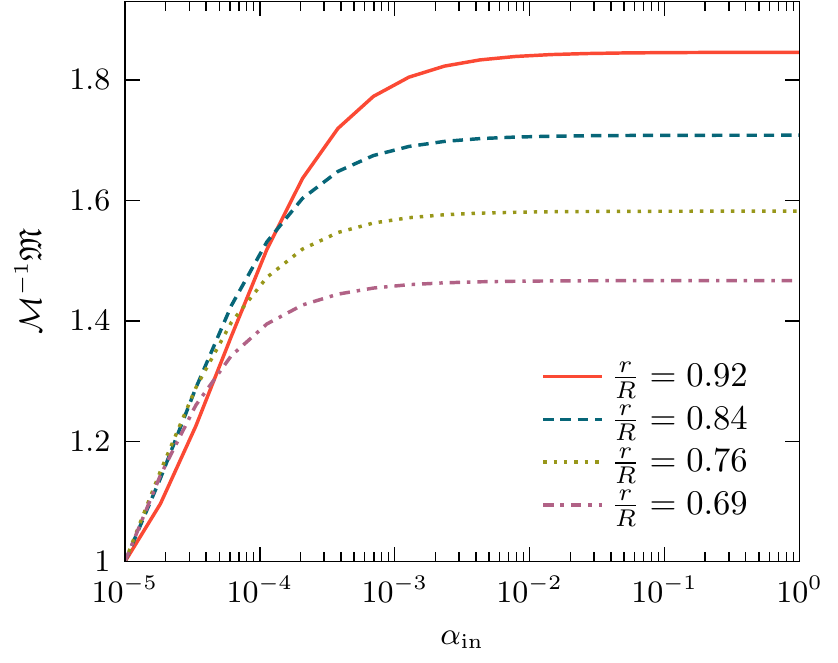}
  \caption{Saturated mean-squared displacement for a thigmotaxis
    process for different values of the ratios of $r/R$. We study the
    dynamics as a function of the normalised parameter
    $\decsymbol_{{i}} \in [0, 1]$ which represents the tendency of the
    walker to remain close to the boundary. When
    $\decsymbol_{{i}} =0$, there are no outer or inner regions, while
    with $\decsymbol_{{i}} = 1$, the walkers never leave the outer
    region once they they get there. The saturation MSD is normalised
    by $\smsd{}$, which is the saturation value when
    $\decsymbol_{{i}} = 0$. Other parameters used are described in in
    the caption of \cref{fig:thigomotaxis_probability}.}
    \label{fig:saturation_msd}
\end{figure}

Given these constraints we study the dynamics as a function of the
$\decsymbol_{i}$. In \cref{fig:thigomotaxis_probability}, we plot the
probability $\rdproptt{\nvec}{\novec}$ for different values of
$t = 500, 750, 1000$ and $\infty$. The walker is initially at the
centre of the domain $\novec = (51, 51)$ and can freely move inside
the inner-region and is able to enter into the outer region without
any resistance. However, once inside the outer region there is a
greater tendency not to leave, due to the high value of
$\decsymbol_{{i}} = 0.95$. We observe this effect when going from
panel (a) to (d). Initially the separation between the inner and outer
regions are barely visible but as time progresses this separation
becomes increasingly clear, culminating with a sharp step at the
steady-state. In panel (e) we plot a cross section of the probability
at $n_2 = 51$ for the times corresponding with panels (a)-(d).

To examine the system further, we plot in \cref{fig:saturation_msd}
the mean-squared displacement (MSD) at steady-state, $\rdsmsd$, as a
function of $\decsymbol_{{i}}$, for four different ratios of inner and
outer regions. The MSD at steady-state is given by
$\rdsmsd = \sum_{\nvec} \ls (n_{_1} - n_{0_1})^2 + (n_{_2} -
n_{0_2})^2 \rs \rdmret^{-1}_{\nvec}$. With the curves normalised to
the case where there are no internal barriers, $\smsd{}$, i.e.  when
$\decsymbol_{{i}} = 0$. We find that as we increase $\decsymbol_{{i}}$
from zero, for small values of $\decsymbol_{{i}}$, $\rdsmsd$ initially
increases logarithmically, while further increases of
$\decsymbol_{{i}}$ causes $\rdsmsd$ to saturate. The value of
saturation is dependent on the ratio of $r/R$: with a high ratio the
outer region is thinner keeping the walker closer to the boundary and
yielding greater saturation value, whereas a smaller ratio results in
a thicker outer region allowing the walker to remain closer to the
initial condition leading to a smaller value of $\rdsmsd$. Note that
the reason for the $r/R = 0.92$ curve not being on top of the others is
due to the discretisation of space when the outer region is very thin.

\section{Two Particle Coalescing Process}
\label{sec:coalescing}
In this final example, we demonstrate the use of our frame to model
certain inert interactions between particles. The interactions we
consider are partial mutual exclusion and reversible binding, both of
which play an important role in coalescing dynamics.
\begin{widetext}

\begin{figure}[H]
  \centering
  \includegraphics{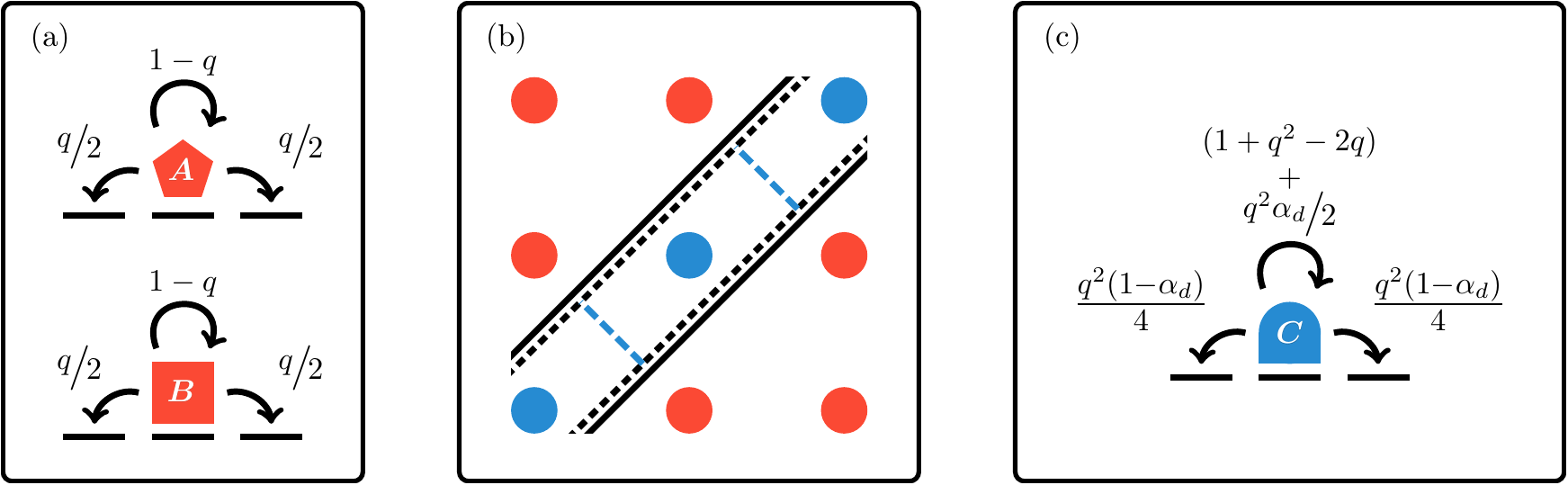}
  \caption{Schematic representation of a two-particle coalescing
    process, modelled as one dimensional interacting random
    walkers. Panel (a) depicts the dynamics of particles $\bm{A}$ and
    $\bm{B}$ where $q \in (0, 1]$ is the probability of moving at each
    time step. The combined dynamics of $\bm{A}$ and $\bm{B}$ can be
    represented via one next-nearest random walker in a 2D
    domain. This abstract domain is depicted in panel (b). The red
    circles represent locations where the two particles are on
    different sites, while the blue circles along the right diagonal
    are locations where they are co-located. In this space the
    interaction of $\bm{A}$ and $\bm{B}$ is modelled with partially
    reflecting barriers. The placement of these barriers are
    illustrated by the solid, dashed, and blue-dashed lines, the
    precise locations and permeability are given in
    \cref{sec:coalescing_defect_placement}. The solid black lines are heterogeneities
    used to model the binding interactions, while the dashed black
    lines are used to control the unbinding interactions. The movement
    of $\bm{C}$ is represented by the 2D random walker moving along
    the diagonal. Its movement is slowed down, relative to $\bm{A}$
    and $\bm{B}$, through the placement of partially reflecting
    barriers along the diagonal depicted by the dashed blue lines. The
    resulting movement dynamics of the complex $\bm{C}$ is shown in
    panel (c), where $\decsymbol_{c} \in [0, 1]$ represents the degree
    with which the movement of the complex $\bm{C}$ is slowed down.}
    \label{fig:coalescing_schematic}
\end{figure}

\end{widetext}

Coalescing processes are ubiquitous in biology and chemistry; they
consist of two or more entities that interact to bind and form a new
one with different movement characteristics. An example of a
coalescing process is the search of a promoter region on DNA by
transcription factors. These movement dynamics alternates between
periods of 3D search in the cytoplasm and periods of restricted search
along the 1D DNA \cite{iwaharakolomeisky2021}. We
use our framework to study a system of relevance to the latter
scenario: a first-passage process of two interacting particle in 1D.

We consider two particles labelled $\bm{A}$ and $\bm{B}$ that move
independently on a 1D lattice with reflecting boundary conditions (see
\cref{fig:coalescing_schematic} for a schematic representation of the
process). Their combined dynamics is described by a two dimensional
next-nearest propagator $\proptt{\nvec}{\novec}$, with
$\novec = (n_{0_1}, n_{0_2})$ and $\nvec = (n_{_1}, n_{_2})$. It
represents the probability that the particle $\bm{A}$ and $\bm{B}$ are
located, respectively, on the site $n_{_1}$ and $n_{_2}$ at time $t$
given that they started, respectively, on $n_{0_1}$, and
$n_{0_2}$. Two particles instantaneously form the complex $\bm{C}$,
namely when they encounter each other, that is when $\nvec = (m, m)$
for $1 \leq m \leq N$.

The interactions between particles is modelled through the placement
of heterogeneities on the combined 2D lattice, yielding three control
parameters, $\decsymbol_{e} \in [0, 1]$, $\decsymbol_{u} \in [0, 1]$,
and $\decsymbol_{c} \in [0, 1]$ (see
\cref{sec:coalescing_defect_placement} for details regarding the
placement of the defects). These parameters are used to constrain,
respectively, the binding events via mutual exclusion of $\bm{A}$ and
$\bm{B}$, the unbinding events of $\bm{C}$ and the mobility of
$\bm{C}$. The parameter $\decsymbol_{u}$ is proportional to the
unbinding probabilities, while $\decsymbol_{e}$ is proportional to the
mutual exclusion probability. When $\decsymbol_{u} = 1$ and
$\decsymbol_{e} = 0$, there is no interaction between the two
particles. The other extreme represents strong interaction: when
$\decsymbol_{e} = 1$ there is mutual exclusion, whereas
$\decsymbol_{u} = 0$ results in a binding that is irreversible.  The
parameter $\decsymbol_{c} \in [0, 1]$ represents the fraction of
the movement probability of complex $\bm{C}$ relative to the movement
probability of the constituent particles $\bm{A}$ and $\bm{B}$. When
$\decsymbol_{c} = 1$ there is no slowing down, while
$\decsymbol_{c} = 0$ results in an immobile $\bm{C}$.

\begin{figure}[ht]
  \centering
  \includegraphics{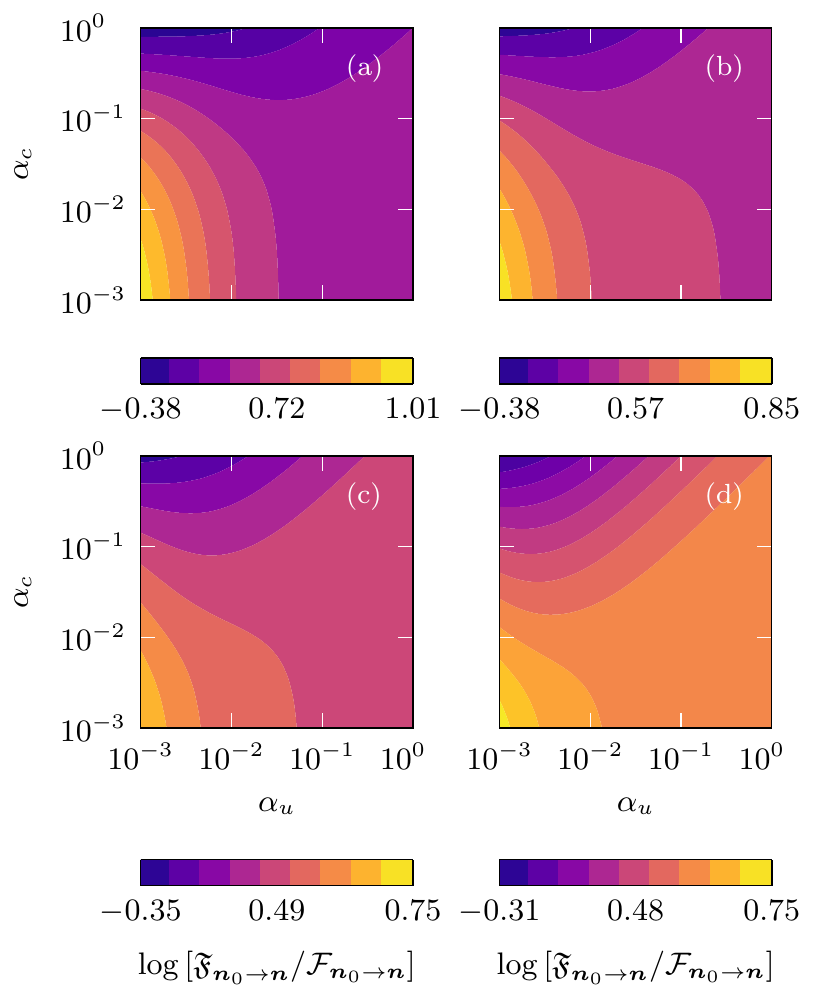}
  \caption{The ratio of the MFPT ($\rdmfptt{\nvec}{\novec}$) of the
    coalescing system compared to the MFPT ($\mfptt{\nvec}{\novec}$)
    of a homogeneous 2D next-nearest neighbour walker as a function of
    the heterogeneity strength parameters (see
    \cref{fig:coalescing_schematic} for detailed a description of the
    parameters involved). The reactive site is located at
    $\nvec = (100, 100)$ and the two particles are initially maximal
    distance away from each other, i.e. $\novec = (1, 100)$ and a
    target location $\nvec = (100, 100)$, on combined 2D domain of
    size $\mvec{N} = (100, 100)$ with diffusion parameter $q =
    2/3$. From panel (a) to (d) we have, respectively, the parameters
    $\decsymbol_{e} = 0, 0.5, 0.75$ and $0.875$.}
  \label{fig:coalescing_mfpt}
\end{figure}

In \cref{fig:coalescing_mfpt} we plot the log ratios of the MFPT,
$\rdmfptt{\nvec}{\novec}$ for both particles to reach a site at the
same time, compared with the 2D homogeneous next-nearest neighbour
analogue, $\mfptt{\nvec}{\novec}$. The latter corresponds with the
case when $\decsymbol_{e}=0$ and $\decsymbol_{u}= \decsymbol_{c} =
1$. The panels (a)-(d) depict $\rdmfptt{\nvec}{\novec}$ for increasing
values of $\decsymbol_{e}$. The smallest ratios are observed in the
upper left quadrant, which corresponds with high cohesiveness of the
complex $\bm{C}$ and with only a slight reduction to its mobility,
given respectively by, low values of $\decsymbol_{u}$ and high values
of $\decsymbol_{c}$. Within this parameter region, once the two
particles bind they rarely separate, consequently the search in 2D
reduces to a search in 1D with fewer sites to explore leading to
smaller \rdmfptt{\nvec}{\novec}.

When there is no exclusion interaction, i.e. panel (a), the dynamics
of a similar model was explored in Ref.~\cite{shinkolomeisky2019}. In
their analysis using asymptotics and simulations, equivalent features
were observed. The most prominent feature of those and our
observations is the minimisation of the MFPT for a slow moving
$\bm{C}$. In this regime, it is more favourable to have an
intermediate unbinding probability, allowing the two particles to
travel independently towards the target before recombining and hitting
the target.

The ability to explore easily the parameter space of the model allows
us to analyse the MFPT for different values of $\decsymbol_{e}$. By
comparing the four panels we observe that as $\decsymbol_{e}$
increases, the overall magnitude of the MFPT ratio decreases. This is
explained by the fact that for small and intermediate values of
$\decsymbol_{e}$ the 2D walker is partially restricted to the upper or
lower triangular regions of the domain, thereby reducing the overall
exploratory space resulting in shorter search times. However, if
$\decsymbol_{e}$ is increased further, i.e. when
$0 \ll \decsymbol_{e} < 1$, the particles will rarely coalesce, and
the MFPT increases. In other words, shorter MFPTs can be achieved by
having particles that mutually exclude one another with some
probability.

\section{Conclusion}
\label{sec:conclusion}

We have introduced an analytical framework to model explicitly any
inert particle-environment interactions.  We have constructed the
discrete Master equation that describes the spatiotemporal dynamics of
diffusing particles in disordered environments by representing the
interactions as perturbed transition dynamics between lattice
sites. To solve this Master equation we have extended the defect
technique to yield the generating function of the propagator in closed
form. Using the propagator, we have derived useful quantities in the
context of transport processes, namely, first-passage, return and exit
probabilities and their respective means. We have also uncovered the
existence of a disorder indifference phenomenon of the mean
first-passage time in quasi 1D systems.

In light of the relevance of our framework to empirical scenarios, we
have chosen three examples to demonstrate its applicability of our
theory. In the first example we consider transdermal drug delivery, an
intercelluar transport process, where we represent the
`brick-and-mortar' structure of the stratum corneum with the placement
of reflecting and partially reflecting barriers. This representation
allows us to study the effect that piercing has on the traversal time
of a drug molecule. In the second example, we have examined the effect
that an animal's thigomotactic response has on the mean squared
displacement at log times. Lastly, in our third example, we have
highlighted the ability of our formalism to study inert interactions
between particles.  We transformed these interactions and the ensuing
dynamics into a single particle moving and interacting with quenched
disorder in a higher dimensional space. The setup allows us to model
analytically the search statistics in a two particle coalescing
process, akin to the search of binding sites on the DNA by multiple
transcription factors.

The strength of our result is in deriving the propagator in the
presence of spatial heterogeneities, $\rdpropzz{\nvec}{\novec}$, as a
function of the homogeneous propagator, i.e. the propagator in the
absence of heterogeneities, $\propzz{\nvec}{\novec}$.  This modularity
allows one to change the movement dynamics by selecting different
forms of $\propzz{\nvec}{\novec}$. In place of the diffusive
propagator one may employ a biased lattice random walk
\cite{sarvaharmangiuggioli2020}, or a walk in different topologies
such as triangular lattices \cite{hughesbook1995,batchelorhenry2002},
Bethe lattices \cite{hughessahimi1982,hughesetal1983} or more
generally a network \cite{masudaetal2017}.

The modularity carries through to the heterogeneous propagator. This
means that in situations where homogeneous space is assumed, one can
relax this assumption and replace the homogeneous propagator,
$\propzz{\nvec}{\novec}$ with the heterogeneous counterpart
$\rdpropzz{\nvec}{\novec}$. We have demonstrated this aspect by
studying the first-passage probability to either of two targets using
results previously derived considering a homogeneous lattice. Further
theoretical exploration could include the analysis of cover time
statistics \cite{nemirovskycoutinho-filho1991,chupeauetal2015},
transmission dynamics
\cite{giuggiolietal2013a,kenkregiuggiolibook2021}, resetting walks
\cite{evansmajumdar2011,palreuveni2017,giuggiolietal2019}, mortal
walks \cite{yusteetal2013}, or random walks with internal degrees of
freedom \cite{montroll1969a}.

Directions for future applications span across spatial and temporal
scales: the role of a building geometry or floor plan on infection
dynamics in hospital wards and supermarkets
\cite{qianetal2009,yingoclery2021,giuggiolisarvaharman2022}; the
prediction of search pattern behaviour of animals in different types
of vegetation cover \cite{cristetal1992,voigtetal2020}; the heat
transfer through layers of skin with differing thermal properties
\cite{simpsonetal2021}; and the influence of topological defects on
the diffusive properties in
crystals~\cite{bauschetal1994,chendeem2003} and territorial
systems~\cite{giuggiolikenkre2014,heiblumroblesgiuggioli2018,sarvaharmanetal2019a}.

We conclude by drawing the reader's attention to the following. As
experimental technologies continue to evolve, observations of the
dynamics of particle-environment interactions are increasing in number
and resolution. The detailed description of the environment that these
technologies bring presents a unique opportunity to rethink modelling
techniques, moving away from macroscopic paradigms to a more
microscopic prescription. We believe that the mathematical framework
we have introduced to quantify the particle-environment interactions
will play a crucial role in connecting the microscopic dynamics to the
macroscopic patterns observed across a vast array of systems.

\begin{acknowledgments}
  SS and LG acknowledge funding from, respectively the Biotechnology
  and Biological Sciences Research Council (BBSRC) Grant
  No.~BB/T012196 and the Engineering and Physical Sciences Research
  Council (EPSRC) Grant No.~S108151-11. We thank Debraj Das and Toby
  Kay for useful discussions.
\end{acknowledgments}

\FloatBarrier
\appendix
\section{Mean First-Passage Statistics}
\label{sec:mfpt}
Using the renewal equation the first-passage probability to a target
is given by the well-known relation
\begin{equation}
  \label{eq:rdfpz_props}
  \begin{aligned}
  \rdfpz_{\novec}(\nvec, z) &=
   \frac{\rdpropz_{\novec}(\nvec, z)}{\rdpropz_{\nvec}(\nvec, z)} \\ &=
\frac {\lb \propz_{\novec}(\nvec, z) - 1 \rb \mdet{\hmat} + \mdet{\hcnmat[0]}}
{\lb \propz_{\nvec}(\nvec, z) - 1 \rb \mdet{\hmat} + \mdet{\hcnmat}},
  \end{aligned}
\end{equation}
where $\hmat$ and $\hcnmat[0]$ are given by, respectively,
Eqs.~\eqref{eq:hmat_mt}, \eqref{eq:hcn0mat_mt} and
\eqref{eq:hcn0mat_mt} with the initial condition being $\nvec$. The
mean of the distribution,
$ \rdmfptt{\nvec}{\novec} = \ld \frac{\mathrm{d}}{\mathrm{d} z}
\rdfpzz{\nvec}{\novec} \rp_{z = 1}$, (see \cref{sup-sec:mret_deriv} of
the Supplementary Materials) is given by
\begin{equation}
  \label{eq:rdmfpt_sup}
\begin{split}
\rdmfpt_{\novec \to \nvec}
= \frac {\mfptt{\nvec}{\novec} \mdet{\hmatd - 1/\mfptt{\nvec}{\novec}\hsmatd[1]}}
{\mdet{\hmatd - \hsmatd[2]}}.
\end{split}
\end{equation}
where  
\begin{align}
  \label{eq:mean_hij_final}
\hmatd_{i, j} &= \frac{\rjq[i]}{\mrett{\vs_i}} \mfpttdif{\vs_i}{\vs_j}{\vps_j} -
                \frac{\ljq[i]}{\mrett{\vps_i}} \mfpttdif{\vps_i}{\vs_j}{\vps_j}
+ \delta_{i, j},\\
  \label{eq:mean_h1ij_final}
\hsmatd[1]_{i, j} &= \lb 
\frac{\rjq[i]}{\mrett{\vs_i}}  \mfpttdif{\vs_i}{\novec}{\nvec} -  
\frac{\ljq[i]}{\mrett{\vps_i}} \mfpttdif{\vps_i}{\novec}{\nvec}  \rb \\ &\times
\mfpttdif{\nvec}{\vs_j}{\vps_j}, \nonumber \\
  \label{eq:mean_h2ij_final}
  \hsmatd[2]_{i, j} &= \lb \frac{\rjq[i]}{\mrett{\vs_i}}   -
                      \frac{\ljq[i]}{\mrett{\vps_i}} 
                      \rb \mfpttdif{\nvec}{\vs_j}{\vps_j} .
\end{align}
If the homogeneous propagator is diffusive with no bias and if the
heterogeneity parameters are symmetric, i.e. $\rjq = \ljq$,
$\hsmatd[2] = 0$ and \cref{eq:rdmfpt_mt} can be simplified further
\begin{equation}
  \label{eq:rdmfpt_sym}
  \rdmfptt{\nvec}{\novec}
  =
  \mfptt{\nvec}{\novec} - 1 
  + 
  \frac
  {
    \mdet{\hmatd - \hsmatd[1]}
  }
  {
    \mdet{\hmatd}
  }.
\end{equation}
\subsection{Mean Return Time}\label{ap:mret}
Through the renewal equation we also have the return probability
relation
\begin{equation}
  \label{eq:rdretz_props}
  \begin{aligned}
  \rdretzz{\nvec} &= 
   1 - \frac{1}{\rdpropz_{\nvec}(\nvec, z)} \\ &= 
\frac{\lb \propzz{\nvec}{\nvec} - 2 \rb \mdet{\hmat} + \mdet{\hcnmat}}
{\lb \propzz{\nvec}{\nvec} - 1 \rb \mdet{\hmat} + \mdet{\hcnmat}}.
  \end{aligned}
\end{equation}
By noticing the identical structures of
\cref{eq:rdfpz_props,eq:rdretz_props}, one can use a similar procedure
to the one used to derive the MFPT (see \cref{sup-sec:mret_deriv} of
the Supplementary Materials) to show the mean return time
(MRT) to be
\begin{equation}
  \label{eq:rdmret_sup}
\rdmrett{\nvec}
=
\frac
{
\mrett{\nvec}
\mdet{\hmatd}
}
{
\mdet{\hmatd - \hsmatd[2]}
}.
\end{equation}



\subsection{Mean Exit Times}\label{ap:mexit_deriv}
The first-exit probability is given by
$\rdexitzz{\novec} = 1 - (1 - z) \rdsurvzz{\novec}$, where
$\rdsurvzz{\novec} $ is the survival probability given by
\begin{equation}
  \label{eq:rdsurv_setup}
  \rdsurvzz{\novec} = \sum_{\nvec}\rdpropzz{\nvec}{\novec}.
\end{equation}
Substituting \cref{eq:rdprop_sol_mt} into \cref{eq:rdsurv_setup} and
evaluating the sum in $\nvec$ and simplifying the summation over $k$
one finds
\begin{equation}
  \rdsurvzz{\novec} = \survzz{\novec} - 1  + \frac{\mdet{\hmat - \hsurvmat}}{\mdet{\hmat}}
  \label{eq:rdsurvzz}
\end{equation}
where $\survzz{\novec} = \sum_{\nvec}\propzz{\nvec}{\novec}$ is the
homogeneous survival probability, and where the elements of $\hmat$
are given in \cref{eq:hmat_mt} and
\begin{equation}
\begin{aligned}
\hsurvmat_{i, j} &= 
\survzz{\initdiff{\vs_j}{\vps_j}}\ls \rjq[i] \propzz{\vs_i}{\novec} \rd \\
& \ld \phantom{\survzz{\initdiff{\vs_j}{\vps_j}}} - \ljq[i] \propzz{\vps_i}{\novec} \rs.
\end{aligned}
\label{eq:hsurv_mat}
\end{equation}
By taking the mean of the first-exit distribution,
i.e.
$\ld \frac{\mathrm{d}}{\mathrm{d} z}\rdexitzz{\novec}\rp_{z = 1}$ 
gives \cref{eq:rdmexit_mt}.
The mathematical details to derive \cref{eq:rdmexit_mt} are given in
\cref{ap:mexit_deriv} while simple expressions of the 1D problem
are given in \cref{sup-sec:expression_1d_suppl} of the Supplementary
materials.


\section{First-passage quantities in 1D systems}
\renewcommand{\thefigure}{B\arabic{figure}}
\setcounter{figure}{0}
\subsection{The MFPT disorder indifference phenomenon}
\label{sec:one_barrier}
\begin{figure}[ht]
  \centering
  \includegraphics{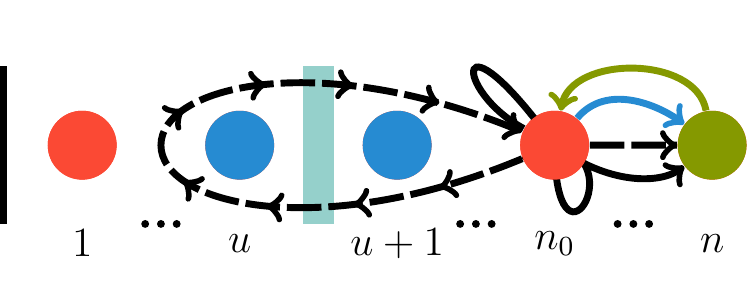}
  \caption{\label{fig:one_barrier_schematic} A schematic
    representation of a one dimensional heterogeneous lattice with a
    reflecting boundary to the left (vertical line) and with a
    permeable barrier between the sites $u$ and $u+1$ represented by
    the shaded rectangle. The first-passage event can be split into
    mutually exclusive events represented by arrows of different
    colours. The blue arrows represent trajectories that never return
    to the initial site, while the black ones represent trajectories
    that return $m$ times before reaching $n$. The green arrow
    represents first-passage trajectories that reach $n_0$ having
    starting at $n$. The solid arrows represent trajectories that are
    unaffected by the presence of the partially reflecting barrier
    between $u$ and $u+1$, while the trajectories that are affected
    are represented by dashed arrows. Note that this schematic depicts the case when $1 \leq u \leq n_0-1$}
\end{figure}

We start with a heterogeneous lattice reflecting boundary between
$n=0$ and $n=1$, and a partially reflecting barrier between $u$ and
$u+1$, with $u < n_0 < n$ as depicted in
\cref{fig:one_barrier_schematic}. The trajectories that contribute to
the first-passage probability can be split into mutually exclusive
sets based on the number of return visits, $m$, to the initial site
$n_0$.

We now formally represent the first-passage probability in terms of a
set of mutually exclusive independent events. Let us define
$\fpt_{n_0}(n, t; m = 0)$ as the first-passage probability to reach
$n$ for the first time at $t$ having started at $n_0$ and having never
returned to the initial site. Clearly, the trajectories that make up
$\fpt_{n_0}(n, t; m=0)$ (coloured blue in
\cref{fig:one_barrier_schematic}), can never be affected by the
presence of the barrier as they never move towards the barrier. The
trajectories that \textit{could} be affected by the presence of the
barrier are those that return at-least once to the initial site before
reaching $n$. The first-passage probability to visit $n$ and having
visited the initial site $m$ times is constructed through the
convolution (dashed trajectories in \cref{fig:one_barrier_schematic})
\begin{equation}
  \newcommand{\tp}[1]{\ensuremath{t_{{#1}}}}
  \label{}
  \begin{aligned}
    \rdfpt_{n_0}(n, t; m) =  &\sum_{\tp{1} = 0}^{t}\cdots\sum_{\tp{m} = 0} ^{\tp{m-1}}\fpt_{n_0}(n, t - \tp{1}; m = 0)  \\
    &\times h_{n_0}(n, \tp{m-1} - \tp{m}) \cdots h_{n_0}(n,  \tp{m}),
  \end{aligned}
\end{equation}
with $t_m \leq t_{m-1} \leq \cdots \leq t_1 \leq t$, and where
\begin{equation}
  \label{}
  \begin{aligned}
    h_{n_0}(n, t) &= 
                    \rdretsymbol(n_0, {t}) \\ &- \sum_{{t'} = 0}^{t}\fpt_{n_0}(n, {t'}; m=0) \fpt_{n}(n_0, {t} - {t'})
  \end{aligned}
\end{equation}
The function $h_{n_0}(n, t)$ represents the probability of returning
without visiting the target and is constructed by considering the
probability of returning to $n_0$ and subtracting those that reach $n$
without returning to $n_0$ at some prior time and subsequently
reaching $n_0$ from $n$. In $z$ domain the relation can be written
more conveniently as
\begin{equation}
  \label{eq:fpz_sys_a}
  \begin{aligned}
\rdfpz_{n_0}(n, z; m) &= \fpz_{n_0}(n, z; m = 0) \ls \rdretzz{n_0} \rd \\ & \ld - \fpz_{n_0}(n, z; m=0) \fpzz{n_0}{n}  \rs^{m}.
  \end{aligned}
\end{equation}
Notice that the only term with the dependence on the barrier on the
RHS of \cref{eq:fpz_sys_a} is $\rdretzz{n_0}$, and in the absence of
the barrier $\rdretzz{n_0}$ reduces to $\retzz{n_0}$. The full
first-passage probability for the system with the barrier, which can
be written as the sum of the mutually exclusive probabilities giving
\begin{equation}
  \label{eq:full_fpz_sys_a}
  \begin{aligned}
\rdfpzz{n}{n_0} &= \fpz_{n_0}(n, z; m = 0)\sum^{\infty}_{m = 0}\ls \rdretzz{n_0} \rd \\ &- \ld \fpz_{n_0}(n, z; m=0) \fpzz{n_0}{n}  \rs^{m}.
  \end{aligned}
\end{equation}
The relation given by \cref{eq:full_fpz_sys_a} is an alternative
method of constructing the first-passage probability i.e. its not one
of the standard approaches which are through the survival probability
or the ratio of propagators in $z$-domain.

To confirm the normalisation of the RHS of \cref{eq:full_fpz_sys_a}
consider the following. By definition $\fpt_{n_0}(n, t; m = 0)$ is not
normalised over $t$, hence, $\fpz_{n_0}(n, z=1; m = 0) = p$ where
$0 < p < 1$. Since all other terms in the RHS of
\cref{eq:full_fpz_sys_a}, namely, $\fpzz{n_0}{n}$ and $\rdretzz{n_0}$
are normalised over time, we find that at $z = 1$ the RHS becomes
$\sum_{m = 0}^{\infty}p(1 - p)^{m} = 1$. Differentiating
\cref{eq:full_fpz_sys_a} with respect to $z$ and taking the limit
$z \to 1$, we obtain the mean first-passage time
\begin{equation}
  \label{eq:rdmret_with_p}
\rdmfptt{n}{n_0} = \frac{\rdmrett{n} - p \mfptt{n_0}{n}}{p}.
\end{equation}
When the barrier is such that
$\jqsymbol_{u, u+1} = \jqsymbol_{u+1, u} = \jqsymbol$ , the mean
return time is equal to the reciprocal of the steady state value, and
$\rdmrett{n_0}$ becomes $\mrett{n_0}$, i.e. the mean return time in
the absence of the barrier.

To find $p$ in \cref{eq:rdmret_with_p} explicitly, we first construct
$\fpz_{n_0}(n, z; m = 0)$ in terms of known quantities using the
approach presented in Ref.~\cite{giuggioli2020} to construct
time-dependent splitting probabilities. We write the two relations by
considering the two splitting separately: the first-passage
probability of reaching the target $n$ and never returning to the
initial condition $\fpt_{n_0}(n, t; m = 0)$; and the first-return
probability to $n_0$ and having never reached the target site $n$. In
time domain they are written via a convolution and are, respectively,
\begin{equation}
  \fpt_{n_0}(n, t; m = 0) = \fpt_{n_0}(n, t) - \sum_{t' = 0}^t  \rdretsymbol(n_0, t; n) \fpt_{n_0}(n, t - t')
\end{equation}
and
\begin{equation}
  \rdretsymbol(n_0, t; n) = \rdretsymbol(n_0, t) - \sum_{t' = 0}^t  \fpt_{n_0}(n, t; m = 0) \fpt_{n_0}(n, t - t'),
\end{equation}
where $\rdretsymbol(n_0, t; n)$ is the probability of returning to the
site $n_0$ at $t$ and having never visited the target $n$. One can
take the $z$-transform and solve for $\fpt_{n_0}(n, z; m = 0)$ and
$\rdretsymbol(n_0, t; n)$ giving, respectively,
\begin{equation}
  \label{eq:fp_z_s}
\fpz_{n_0}(n, z; m = 0) = \frac{\fpzz{n}{n_0} - \rdretzsymbol(n_0, z)\fpzz{n}{n_0}}{1 - \fpzz{n}{n_0}\fpzz{n_0}{n}}
\end{equation}
and
\begin{equation}
  \label{eq:ret_z_s}
\rdretzsymbol(n_0, z; n) =  \frac{\rdretzsymbol(n_0, z) - \fpzz{n}{n_0}\fpzz{n_0}{n}}{1 - \fpzz{n}{n_0}\fpzz{n_0}{n}}.
\end{equation}
Evaluating \cref{eq:fp_z_s,eq:ret_z_s} at $z = 1$ gives, respectively,
the fraction of all the first-passage trajectories that reach the
target without returning to $n_0$ and the fraction of all trajectories
that return to $n_0$ without ever reaching $n$. Using de
L'H\^{o}pital's rule once in \cref{eq:fp_z_s} we find
\begin{equation}
  \label{eq:p_simp}
  \begin{aligned}
    p &= \frac{\rdmrett{n_0}}{\mfptt{n}{n_0} + \mfptt{n_0}{n}}.
  \end{aligned}
\end{equation}
Inserting \cref{eq:p_simp} in \cref{eq:rdmret_with_p} one finds that
$\rdmfptt{n}{n_0} = \mfptt{n}{n_0}$.
\subsection{MFPT linear dependence on disorder location}
\label{sec:linear_dep_disorder}
To understand the linear dependence in $u$, with $n_0 \leq u < n$,
present in \cref{eq:rdmfpt_ref_diff_1d}, we consider building up first
passage probability by convolution in time to go from $n_0$ to $u$
first, then from $u$ to $u+1$ and then from $u+1$ to $n$. In $z$
domain one has
\begin{equation}
  \label{eq:fpz_conv}
\rdfpzz{n}{n_0} = \fpzz{u}{n_0} \rdfpzz{u+1}{u} \rdfpzz{n}{u+1},
\end{equation}
where the first term on the RHS has no dependence on the barrier as it
is after the absorbing site $u$, while the other two terms are
dependent on the barrier. Computing the mean of \cref{eq:fpz_conv}, we obtain
\begin{equation}
  \label{eq:mfpt_1d_sum}
\rdmfptt{n}{n_0} = \mfptt{u}{n_0}  + \rdmfptt{u+1}{u}  + \mfptt{n}{u+1},
\end{equation}
where we have substituted $\rdmfptt{n}{u+1} = \mfptt{n}{u+1}$ using
the justification presented in the previous section. By using the
relation $\mfptt{n}{n_0} = \mfptt{s}{n_0} + \mfptt{n}{s}$ with
$n_0 < s < n$, one can rewrite \cref{eq:mfpt_1d_sum} to give
\begin{equation}
  \label{}
\rdmfptt{n}{n_0} = \mfptt{n}{n_0}  + \rdmfptt{u+1}{u}  - \mfptt{u+1}{u}.
\end{equation}
In the diffusive case, the MFPT to a neighbouring site,
$\mfptt{u+1}{u}$, is always proportional to twice the distance between
$u$ and the reflecting boundary to the left, i.e.
\begin{equation}
  \label{eq:hom_mfpt_neigh}
\mfptt{u+1}{u} = \frac{2}{q}(u - s + 1),
\end{equation}
with $s \leq u$ being the position of the reflecting boundary. By
simplifying the general MFPT given in \cref{eq:rdmfpt_mt}, we find
\begin{equation}
  \label{eq:het_mfpt_neigh}
\rdmfptt{u+1}{u} = \frac{2}{q - 2 \jqsymbol} \lb u - s + 1\rb,
\end{equation}
which is analogous to \cref{eq:hom_mfpt_neigh} but with the
multiplicative (time rescale) factor increased to
$2 (q - 2 \jqsymbol)^{-1}$.  Letting $s = 1$ we obtain
\cref{eq:rdmfpt_ref_diff_1d}.

\section{Placement of Defects and Parameter Choice of the Modelling
  Applications}
\subsection{Thigmotaxis}
\label{sec:thigmotaxis_defects}
Two sets of defects must be placed, one set along a circle radius $R$
to create a (circular) reflecting domain, while the second is used to
divide this domain into two different regions (see
\cref{sec:thigmotaxis}) and is placed along a circle of radius $r$. To
place defects on either circle one must first know which sites are
within which circle. To determine this we use the Euclidean distance
as a heuristic, with the site $\nvec = (n_1, n_2)$ being part of the
circular domain if and only if $h(n_1, n_2) \leq R$, where
$h(n_1, n_2) = \ls {(n_1 - R - 1)}^2 + {(n_2 - R - 1)} \rs^{1/2}$ with
the size of the bounding square domain given by
$\mvec{N} = (2 R + 1, 2 R + 1)$. Similarly, a site is part of the
inner region if and only if $h(n_1, n_2) \leq r$, while the outer
region is given by $r < h(n_1, n_2) \leq R$. Given these site
partitions, one can define two sets of defects, $S_{{d}}$, and
$S_{{i}}$ describing, respectively, the impenetrable barriers to
restrict the walker to a circular domain, and partially-reflecting
inner barriers. In both cases $\vs$ represents sites inside the circle
of defects while $\vps$ represents sites outside. For all
$\lc \vs, \vps \rc \in S_{{i}}$ we have $\rjq = \matjump{\vps}{\vs}$
and $\ljq$ is irrelevant as the walker initially starts inside the
circular domain. For all $\lc \vs, \vps \rc \in S_{{d}}$, we let
$\rjq = 0$ providing no resistance for the walker to enter the
outer-region and $\ljq = \decsymbol_{{i}}\matjump{\vs}{\vps}$ with
$\decsymbol_{{i}} \in [0, 1]$.

\subsection{Two particle coalescing process}
\label{sec:coalescing_defect_placement}
The interactions that need to modelled are binding and unbinding.
Binding can occur via two distinct events. The first is when two
particles are located on neighbouring sites with $\nvec = (m+1, m)$
and at the following time step, one of the particles remains at the
same site while the second particle jumps onto the site occupied by
the first resulting in $\nvec = (m+1, m+1)$ or $\nvec = (m, m)$. The
second possible event occurs when the two particles are located two
sites apart i.e.  $\nvec = (m-1, m+1)$ and at the following time step
they both jump towards each other landing on $\nvec = (m, m)$. The
reverse of these two events gives rise to unbinding of the complex
$\bm{C}$. These transitions can be modified by placing paired defects
of the forms: $\vs = (m, m )$, $\vps = (m+1, m)$, and $\vs = (m, m )$,
$\vps = (m, m+1)$ for $1 \leq m \leq N-1$ with
$\rjq = \frac{q}{2}(1 -q) (1 - \decsymbol_{u})$,
$\ljq = \frac{q}{2}(1 -q)\decsymbol_{e}$; $\vs = (m, m )$ ,
$\vps = (m, m-1)$ and $\vs = (m, m )$ , $\vps = (m-1, m)$ for
$2 \leq m \leq N$ with $\rjq = \frac{q}{2}(1 -q) (1 - \decsymbol_{u})$,
$\ljq = \frac{q}{2}(1 -q)\decsymbol_{e}$; $\vs = (m, m)$,
$\vps = (m\mp 1, m \pm 1)$ for $2 \leq m \leq N-1$ with
$\rjq = \frac{q^2}{4}(1- \decsymbol_{u})$,
$\ljq = \frac{q^2}{4}\decsymbol_{e}$.

Intuitively, the movement of the coalesced is slowed as it more
massive. To encode this detail we interpret jumps along the leading
diagonal $\nvec = (m, m)$ for all $1 \leq m \leq N$ as the jumps made
by the coalesced particle $\bm{C}$, and we slow its movement by
placing paired defects of the form $\vs = (m, m)$, $\vps = (m+1, m+1)$
for all $ 1 \leq m \leq N-1$ with
$\rjq = \ljq = \frac{q^2}{4} (1 - \decsymbol_{c})$.

\bibliography{library1.bib,library2.bib,library3.bib}

\makeatletter\@input{xx.tex}\makeatother
\end{document}


\preprint{APS/123-QED}
\title{Supplementary Materials for the article
                  ``Particle-Environment Interactions In Arbitrary
                  Dimensions: A Unifying Analytic Framework To Model
                  Diffusion With Inert Spatial Heterogeneities''}

\author{Seeralan Sarvaharman}
 \email{Email: s.sarvaharman@bristol.ac.uk}
\affiliation{Department of Engineering Mathematics, University of Bristol, BS8 1UB, UK}
\author{Luca Giuggioli}%
 \email{Email: Luca.Giuggioli@bristol.ac.uk}
\affiliation{Department of Engineering Mathematics, University of Bristol, BS8 1UB, UK}
\affiliation{Bristol Centre for Complexity Sciences, University of Bristol, BS8 1UB, UK}


\maketitle


\onecolumngrid
\tableofcontents

\section{Derivation of the Heterogeneous Propagator}~\label{sec:refl_defect_deriv}

We consider a collection of heterogeneous connections as described in
\cref{mt-sec:heterogenous_propagator} of the main text, given by a set
of $M$ paired defects,
$S = \lc \lc \vs_1, \vps_1 \rc, \cdots, \lc \vs_M, \vps_M
\rc\rc$. When not on a defective site, that is when
$\nvec \neq \vs_k, \vps_k$ for any $k$, the dynamics are given by
\begin{equation}
  \begin{aligned}
  \rdpropt(\nvec, t + 1) &= \sum_{\mvec{m}}\matjump{\nvec}{\mvec{m}}\, \rdpropt(\mvec{m},  t), \quad
  \nvec \neq \vs, \vps, \ \forall \lc \vs, \vps \rc \in S,
  \end{aligned}
\label{eq:rdprop_setup_off_defect}
\end{equation}
with $\mmat{A}_{\nvec, \mvec{m}}$ representing the transition
probability from site $\mvec{m}$ to site $\nvec$. When, instead, on
any of the paired defective sites the dynamics are given by
\begin{equation}
\begin{aligned}
\rdpropt(\vs, t + 1) &= \sum_{\mvec{m}}\matjump{\nvec}{\mvec{m}}\, \rdpropt(\mvec{m},  t)  
+ \rjq \rdpropt(\vs, t)
- \ljq \rdpropt(\vps, t) ,
\end{aligned}
\label{eq:rdprop_setup_left_defect}
\end{equation}
and 
\begin{equation}
\begin{aligned}
\rdpropt(\vps, t + 1) &= \sum_{\mvec{m}}\matjump{\nvec}{\mvec{m}}\, \rdpropt(\mvec{m},  t) 
- \rjq \rdpropt(\vs, t)
+ \ljq \rdpropt(\vps, t) ,
\end{aligned}
\label{eq:rdprop_setup_right_defect}
\end{equation}
where the bounds on $\rjq$ and $\ljq$ are given by
\cref{mt-eq:lam_const_1,mt-eq:lam_const_2} in the main text.
Combining
\cref{eq:rdprop_setup_off_defect,eq:rdprop_setup_left_defect,eq:rdprop_setup_right_defect}
into a single equation and summing over all pairs in $S$ gives the
Master equation with defects
\begin{equation}
\begin{aligned}
  &\rdpropt(\nvec, t + 1) = \sum_{\mvec{m}}\mmat{A}_{\nvec, \mvec{m}} \, \rdpropt(\nvec,  t)  
  + \sum^{M}_{k = 1} \deltadif{\nvec}{\vs_k}{\vps_k} 
  \ls \big. \rjq[k] \rdpropt\lb\vs_k, t\rb -\ljq[k] \rdpropt\lb\vps_k, t\rb \rs,
\end{aligned}
\label{eq:rdprop_setup_combined_defect}
\end{equation}
where $\deltadif{\nvec}{\vs}{\vps} =
\deltadifful{\nvec}{\vs}{\vps}$. Taking the $z$ transform of
\cref{eq:rdprop_setup_combined_defect} we find
\begin{equation}
\label{eq:rdprop_setup_combined_defect_z}
\begin{aligned}
  &\rdpropz(\nvec, z) - \rdpropt(\nvec, 0) =
  \sum_{\mvec{m}}\mmat{A}_{\nvec, \mvec{m}}\, \rdpropz(\nvec,  z) 
  + z\sum_{k = 1}^{M}\deltadif{\nvec}{\vs_k}{\vps_k}
\ls \big. \rjq[k] \rdpropz(\vs_k, z) - \ljq[k] \rdpropz(\vps_{k}, z) \rs.
\end{aligned}
\end{equation}
Solving first the homogeneous difference equation i.e. \cref{eq:rdprop_setup_off_defect} 
to get (in the absence of defects)
\begin{equation}
\rdpropz(\nvec, z) 
= 
 \sum_{\mvec{m}} \propz_{\mvec{m}}(\nvec, z) \rdpropt(\mvec{m}, 0) 
\end{equation}
where $\propz_{\mvec{m}}(\nvec, z)$ is the propagator of the
homogeneous problem (e.g. see Eq.~(23) of Ref.~\cite{giuggioli2020} or
Eq.~(33) of Ref.~\cite{sarvaharmangiuggioli2020}), followed by a
convolution (in time and space) with the inhomogeneous term in
\cref{eq:rdprop_setup_combined_defect_z} yields the formal solution
\begin{equation}
\begin{aligned}
&\rdpropz(\nvec, z) = 
 \sum_{\mvec{m}} \propz_{\mvec{m}}(\nvec, z) \rdpropt(\mvec{m}, 0)  
+ z\sum^{M}_{k = 1} \propzzdif{\nvec}{\vs_k}{\vps_k}  
 \ls \rjq[k] \rdpropz(\vs_k, z) - \ljq[k] \rdpropz(\vps_k, z) \rs.
\end{aligned}
\label{eq:rdprop_defect_formal_sol_z}
\end{equation}
When the initial condition is localised, i.e.
$\rdpropt(\nvec, 0) = \delta_{\nvec, \novec}$, we have the formal
propagator
\begin{equation}
\begin{aligned}
&\rdpropzz{\nvec}{\novec} = 
\propz_{\novec}(\nvec, z)
+ z\sum_{k = 1}^{M}
\propzzdif{\nvec}{\vs_k}{\vps_k} 
 \ls \rjq[k] \rdpropz_{\novec}(\vs_k, z) 
- \ljq[k] \rdpropz_{\novec}(\vps_k, z) \rs.
\end{aligned}
\label{eq:rdprop_defect_formal_sol_z_with_init}
\end{equation}
In order to find $\rdpropz_{\novec}(\nvec, z)$ in terms of the known
propagator $\propz_{\novec}(\nvec, z)$ we first create simultaneous
equations for each pair of defects in terms of the differences
$\rjq[k]\rdpropz_{\novec}(\vs_k, z) - \ljq[k]\rdpropz_{\novec}(\vps_k,
z)$ giving
\begin{equation}
\label{eq:rdprop_defect_formal_sol_z_at_defect}
\begin{aligned}
&\rjq[k]\rdpropz_{\novec}(\vs_k, z) -
\ljq[k]\rdpropz_{\novec}(\vps_k, z) = 
\rjq[k]\propzz{\vs_k}{\novec} - 
\ljq[k]\propzz{\vps_k}{\novec}  \\ &+
 z\sum_{\ell = 1}^{M}
 \ls \bigg.
 \rjq[k]
 \propzzdif{\vs_k}{\vs_{\ell}}{\vps_{\ell}}
 -  
 \ljq[k]
 \propzzdif{\vps_k}{\vs_{\ell}}{\vps_{\ell}}
 \rs
\ls \bigg. \rjq[\ell]\rdpropz_{\novec}(\vs_\ell, z) - \ljq[\ell]\rdpropz_{\novec}(\vps_\ell, z) \rs,
\end{aligned}
\end{equation}
whose solution via Cramer's rule, is given by
\begin{equation}
\begin{aligned}
&\rjq[k]\rdpropz_{\novec}(\vs_k, z) -
\ljq[k]\rdpropz_{\novec}(\vps_k z)
= -\frac{1}{z}\frac{\mdet{\hkmat}}{\mdet{\hmat}},
\label{eq:rdprop_cramers}
\end{aligned}
\end{equation}
with $\hmat$ defined in \cref{mt-eq:hmat_mt} and where $\hkmat$ is
the same as $\hmat$, but with the $k^{\mathrm{th}}$ column replaced
by
\begin{equation}
  \begin{aligned}
    &\ls \rjq[1] \propzz{\vs_1}{\nvec_{_0}} - \ljq[1]\propzz{\vps_1}{\nvec_{_0}}, \cdots,
 \rjq[M] \propzz{\vs_M}{\nvec_{_0}} - \ljq[M]\propzz{\vps_M}{\nvec_{_0}} \rs^{\trans}.
  \end{aligned}
\end{equation}
Using \cref{eq:rdprop_cramers},
\cref{eq:rdprop_defect_formal_sol_z_with_init} becomes
\begin{equation}
  \label{eq:rdprop_sol_full}
  \rdpropz_{\novec}(\nvec, z) =
\propz_{\novec}(\nvec, z) 
-\sum_{k = 1}^{M} \propzzdif{\nvec}{\vs_k}{\vps_k}
\frac{\mdet{\hkmat}}{\mdet{\hmat}}.
\end{equation}
The summation in \cref{eq:rdprop_sol_full} can be carried out
explicitly giving
\begin{equation}
  \rdpropzz{\nvec}{\novec} =
  \propzz{\nvec}{\novec} - 1 + \frac{\mdet{\hmat - \hsmat[0]}}{\mdet{\hmat}}.
  \label{eq:summed_defect_with_hs}
\end{equation}
where,
\begin{equation}
\begin{aligned}
\hsmat[0]_{i, j} = 
\propzzdif{\nvec}{\vs_i}{\vps_i}
& \ls   \big. \rjq[i]\propzz{\vs_i}{\novec} 
 - \ljq[i] \propzz{\vps_i}{\novec} \rs 
\end{aligned}
\end{equation}
and calling $\hcnmat[0] = \hmat - \hsmat[0]$ gives the solution
presented in \cref{mt-eq:rdprop_sol_mt}.



\subsection{Sticky and Slippery Heterogeneities}
\label{sec:sticky_slippery_prop}
\renewcommand{\thefigure}{A\arabic{figure}}
\setcounter{figure}{0}
We start with set of defective sites
$S' = \lc \ws_1, \cdots, \ws_L \rc$ and use the notation $\wsl[i]$ and
$\wsr[i]$ representing, respectively, the left and right neighbours of
$\mvec{w}_i$ in the $s^{\mathrm{th}}$-dimension, given
$\ws = (w_1, \cdots, w_d)$,
$\wsr = (w_1, \cdots w_{s} + 1, \cdots , w_d)$ and
$\wsl = (w_1, \cdots w_{s} - 1, \cdots , w_d)$, with $d$ the lattice
dimension. A schematic representation of the jump probabilities on a
defective site, $\ws$ is given in \cref{fig:sticky_defect_schematic}
\begin{figure}[ht]
  \centering
  \pgfmathsetmacro\siteRad{0.41} 
  \includegraphics{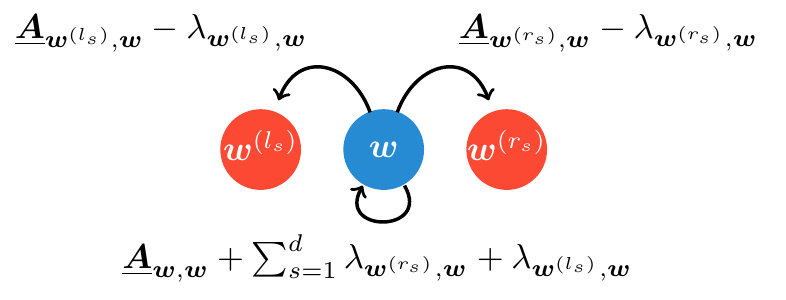}
  \caption{\label{fig:sticky_defect_schematic}A schematic
    representation showing the modified transition probabilities of a
    sticky (or slippery) heterogeneity.  We highlight only the
    dynamics in the $s^{\mathrm{th}}$ dimension, but the same is
    present for the other dimensions.}
\end{figure}, from which it is clear that to ensure positive
probabilities one must have
$\rjq[][\wsr][\ws] \leq \matjump{\wsr}{\ws}$ and
$\ljq[][\wsl][\ws] \leq \matjump{\wsl}{\ws}$ for all
$s = 1, \cdots, d$, and
$0 \leq \matjump{\ws}{\ws} + \sum_{s =
  1}^{d}\rjq[][\wsr][\ws]+\ljq[][\wsl][\ws]$ for all $\ws \in S'$.
These conditions are a recast of the one given by
\cref{mt-eq:lam_const_1,mt-eq:lam_const_2} in the main text.

The full dynamics is
described by the Master equation
\begin{equation}
  \label{eq:rdprop_sticky_full_master}
  \begin{aligned}
\rdproptt[t+1]{\nvec}{} &= \sum_{\mvec{m}} \matjump{\nvec}{\mvec{m}} \rdproptt{\mvec{m}}{} + 
\sum_{k = 1}^{L} \rdproptt{\ws_k}{}
\lc \sum_{s = 1}^{d} 
\rjq[][{\wsr[k]}][{\ws_{k}}]
\deltadif{\nvec}{\ws_k}{{\wsr[k]}} 
 +  \ljq[][{\wsl[k]}][{\ws_k}]
\deltadif{\nvec}{\ws_k}{{\wsl[k]}} \rc.
  \end{aligned}
\end{equation}
Using a localised initial
condition $\rdproptt[0]{\nvec}{} = \delta_{\nvec, \novec}$ and
proceeding as before by solving the homogeneous dynamics and
convolution (in time and space) with the non-homogeneous part of
\cref{eq:rdprop_sticky_full_master} gives the formal solution
\begin{equation}
  \label{eq:rdprop_sticky_formal_sol}
\rdpropzz{\nvec}{\novec} = \propzz{\nvec}{\novec} + 
z\sum_{k = 1}^{M} \rdpropzz{\ws_k}{\novec}
  \sypropzz{\nvec}{\ws_k},
\end{equation}
where
\begin{equation}
  \begin{aligned}
  \sypropzz{\nvec}{\ws} &= 
\sum_{s = 1}^{d} 
\rjq[][{\wsr}][{\ws}]
\propzzdif{\nvec}{\ws}{\wsr} + 
\ljq[][{\wsl}][{\ws}]
\propzzdif{\nvec}{\ws}{\wsl}.
  \end{aligned}
\label{eq:omega_ap}
\end{equation}
This formal solution is as special case of
\cref{eq:rdprop_defect_formal_sol_z_with_init} where $M = 2Ld$ where
each of the sites in $S'$ bears two paired defects for each of $d$
dimensions. However, by noticing that the incoming connections of the
sticky sites are left unmodified,
i.e.  $\ljq[][\ws][\wsl] = \rjq[][\ws][\wsr] = 0$ one can to simplify
\cref{eq:rdprop_defect_formal_sol_z_with_init} to
\cref{eq:rdprop_sticky_formal_sol} thereby reducing the number unknowns
by a factor of $2d$.

To find the full solution we let $\nvec = \ws_k$ and solve the
simultaneous equations
\begin{equation}
  \label{eq:rdprop_sticky_formal_sol_sub_k}
\begin{aligned}
  \rdpropzz{\ws_k}{\novec}
= \propzz{\nvec}{\novec} +
z\sum_{\ell = 1}^{M} \rdpropzz{\ws_\ell}{\novec}
  \sypropzz{\ws_k}{\ws_{\ell}},
\end{aligned}
\end{equation}
with $k = 1, \cdots, M$ to get
\begin{equation}
  \label{eq:rdprop_sticky_sol_k}
\rdpropzz{\ws_k}{\novec} = -\frac{1}{z} \frac{\mdet{\hkmat}}{\mdet{\hmat}}
\end{equation}
where, in this case, the matrix $\hmat$ is simplified to
\begin{equation}
  \label{eq:hmat_sticky}
    \hmat_{i, j} = \sypropzz{\ws_i}{\ws_j} - \frac{1}{z}\delta_{i, j} 
\end{equation}
and $\hkmat$ is the same as $\hmat$ but with the $k^{\mathrm{th}}$
column replaced by
$\ls \propzz{\ws_1}{\novec}, \cdots, \propzz{\ws_M}{\novec}
\rs^{\trans}$. Substituting \cref{eq:rdprop_sticky_sol_k} into
\cref{eq:rdprop_sticky_formal_sol} and summing over $k$ gives the full
solution in \cref{mt-eq:rdprop_sol_mt}  where
\begin{align}
  \label{eq:hcn0mat_sticky}
    \hcnmat[0]_{i, j} &= \hmat_{i, j} - \sypropzz{\nvec}{\ws_j} \propzz{\mvec{w}_i}{\novec}.
  \end{align}

\section{Derivations of first-passage statistics in the presence
of heterogeneities}
\label{sec:mfpt_deriv}
\subsection{Mean first-passage time with arbitrary type and number of heterogeneities}
From the renewal equation, the generating function of the
first-passage probability from $\novec$ to $\nvec$ ($\nvec \neq \novec$) is given by
\begin{equation}
  \label{eq:rdfpz_props}
  \rdfpz_{\novec}(\nvec, z) =
   \frac{\rdpropz_{\novec}(\nvec, z)}{\rdpropz_{\nvec}(\nvec, z)} = 
\frac {\lb \propz_{\novec}(\nvec, z) - 1 \rb \mdet{\hmat} + \mdet{\hmat - \hsmat[0]}}
{\lb \propz_{\nvec}(\nvec, z) - 1 \rb \mdet{\hmat} + \mdet{\hmat - \hsmat}},
\end{equation}
where we have called $\hcmat[0] = \hmat - \hsmat[0]$ with $\hmat$ and
$\hcnmat[0]$ defined, respectively, in \cref{mt-eq:hmat_mt,mt-eq:hcn0mat_mt} of the main text. Note that the matrix
$\hsmatfull{\mvec{n}}{\mvec{m}}$ can be written in the form
$\mvec{a} \mvec{b}^{\trans}$, where $\mvec{a}$ and $\mvec{b}$ are
column vectors with elements
$\mvec{a}_i = \rjq[i] \propzz{\vs_i}{\mvec{m}} - \ljq[i]
\propzz{\vps_i}{\mvec{m}}$ and
$\mvec{b}_i = \propzzdif{\nvec}{\vs_i}{\vps_i}$. We will exploit this
property in the coming steps. Dividing both the numerator and
denominator of \cref{eq:rdfpz_props} by $\propz_{\nvec}(\nvec, z)$
gives
\begin{equation}
  \rdfpz_{\novec}(\nvec, z) =
   \frac {\lb \fpzz{\nvec}{\novec} - 1/\propzz{\nvec}{\nvec} \rb \mdet{\hmat}
     + 1/\propzz{\nvec}{\nvec}\mdet{\hmat - \hsmat[0]}}
   { \lb \bigg. 1 - 1/\propzz{\nvec}{\nvec} \rb \mdet{\hmat} +
     1/\propzz{\nvec}{\nvec}\mdet{\hmat - \hsmat}}.
\end{equation}
Using the property
\begin{equation}
  \label{eq:det_prop}
\decsymbol \mdet{\mmat{A} - \mmat{B}} = \mdet{\mmat{A} - \decsymbol \mmat{B}} - (1 - \decsymbol) \mdet{\mmat{A}}
\end{equation}
when $\alpha$ is a scalar and $\mmat{B} = \mvec{a}\mvec{b}^{\trans}$
with $\mvec{a}$ and $\mvec{b}$ column vectors of appropriate size, we
rewrite
\begin{equation}
  \rdfpz_{\novec}(\nvec, z) =
   \frac {\lb \fpzz{\nvec}{\novec} - 1 \rb \mdet{\hmat}
     + \mdet{\hmat - 1/\propzz{\nvec}{\nvec}\hsmat[0]}
   }{\mdet{\hmat - 1/\propzz{\nvec}{\nvec}\hsmat}}.
\end{equation}
Dividing through by
$ \prod_{k = 1}^{M} \propz_{\vs_k}(\vs_k, z) \propz_{\vps_k}(\vps_k,
z)$, where $M$ is the total number of paired defects one finds
\begin{equation}
  \label{eq:rdfpz_fpz}
 \rdfpz_{\novec}(\nvec, z) =
\frac {\lb \fpzz{\nvec}{\novec}   - 1\rb \mdet{\hmatf} + \mdet{\hcnmatf[0]}}
{\mdet{\hcnmatf}}.
\end{equation}
where the elements of the matrices $\hmatf$, $\hcnmatf[0]$ and
$\hcnmatf$ are given in terms of first-passage and return
probabilities
\begin{align}
  \label{eq:hmat_fpz}
  \hmatf_{i, j} &=
  \rjq[i] \ls 1 - \retzz{\vps_i} \rs\fpzzdif{\vs_i}{\vs_j}{\vps_j} -
  \ljq[i] \ls 1 - \retzz{\vs_i}  \rs\fpzzdif{\vps_i}{\vs_j}{\vps_j} \nonumber \\
  &- \delta_{i, j}z^{-1}\ls 1 - \retzz{\vps_i} \rs \ls 1 - \retzz{\vs_i} \rs, \\
  \label{eq:hcn0mat_fpz}
   \hcnmatf[0]_{i, j} &= \hmatf_{i, j} -
   \fpzzdif{\nvec}{\vs_j}{\vps_j}
  \lc \big.
 \rjq[i]\ls 1 - \retzz{\vps_i} \rs\fpzz{\vs_i}{\novec}
- \ljq[i]\ls 1 -\retzz{\vs_i} \rs \fpzz{\vps_i}{\novec}
 \rc, \\
  \label{eq:hcnmat_fpz}
   \hcnmatf_{i, j} &= \hmatf_{i, j} -
   \fpzzdif{\nvec}{\vs_j}{\vps_j}
  \lc \big.
 \rjq[i]\ls 1 - \retzz{\vps_i} \rs\fpzz{\vs_i}{\nvec}
- \ljq[i]\ls 1 -\retzz{\vs_i} \rs \fpzz{\vps_i}{\nvec}
 \rc.
\end{align}
The mean first-passage time is then given by
\begin{equation}
  \label{eq:mfpz_n_d}
\ld \diffz{\rdfpzz{\nvec}{\novec}} \rp_{z = 1} = 
\ld  \frac{\lb \diffz[]{\mathbb{N}} \rb\mathbb{D}  - \lb \diffz[]{\mathbb{D}} \rb \mathbb{N}}{\mathbb{D}^2} \rp_{z \to 1},
\end{equation}
where $\diffz[k]{f}$ is the $k^{\mathrm{th}}$ derivative of $f$ with
respect to $z$,
$\mathbb{N} = \lb \fpzz{\nvec}{\novec} - 1\rb \mdet{\hmatf} +
\mdet{\hcnmatf[0]}$ and $\mathbb{D} =
\mdet{\hcnmatf}$. When $z \to 1$,
$\mdet{\hmatf}, \mdet{\hcnmatf[0]}$ and $\mdet{\hcnmatf}$ all reduce to
zero and it becomes necessary to use de L'Hôpital's rule. In order to
proceed, it helps to consider the $k^{\mathrm{th}}$ derivative
of a determinant of a matrix with size $M\times M$ given by
\begin{equation}
  \label{eq:det_deriv}
\diffz[k]{\mdet{\mmat{A}}}
=
\sum_{k_1 + \cdots + k_M =k }
\frac{k!}{k_1! \cdots k_M!}
\mdet{\begin{matrix}
 \diffz[k_1]{\mmat{A}_{1,1}}  & \diffz[k_2]{\mmat{A}_{1,2}} & \cdots & \diffz[k_M]{\mmat{A}_{1,M}} \\
 \diffz[k_1]{\mmat{A}_{1,2}}  & \diffz[k_2]{\mmat{A}_{2,2}} & \cdots & \diffz[k_M]{\mmat{A}_{2,M}} \\
 \vdots                      & \vdots                     & \ddots &  \vdots \\
 \diffz[k_1]{\mmat{A}_{M, 1}} & \diffz[k_2]{\mmat{A}_{i,j}} & \cdots & \diffz[k_M]{\mmat{A}_{M,M}}
\end{matrix}}.
\end{equation}
From \cref{eq:det_deriv} and the expressions in
\cref{eq:hmat_fpz,eq:hcn0mat_fpz,eq:hcnmat_fpz} it becomes clear that
each of the columns in the matrices must be differentiated at-least
twice to give a non zero determinant when $z \to 1$. The determinant
must therefore be differentiated $2M$-times leading to de L'Hôpital's
rule being used $4M$-times in \cref{eq:mfpz_n_d}.  Expanding the
denominator using Leibniz general rule gives
\begin{equation}
  \label{}
  \diffz[4M]{\mathbb{D}^2}
  =
\sum_{k = 0}^{4M} 
\begin{pmatrix}
 4M \\ 
 k
\end{pmatrix}
\lb \diffz[k]{\mathbb{D}} \rb
\lb \diffz[4M-k]{\mathbb{D}} \rb 
= 
\begin{pmatrix}
 4M \\ 
 2M
\end{pmatrix}
\lb \diffz[2M]{\mathbb{D}} \rb^{2},
\end{equation}
where the only non-zero term is when $k = 2M$, expanding the first
term in the numerator in \cref{eq:mfpz_n_d} yields
\begin{equation}
\begin{aligned}
  \label{}
  \diffz[4M]{\ls \lb  \diffz \mathbb{N} \rb \mathbb{D} \rs} 
  &=
\sum_{k = 0}^{4M} 
\begin{pmatrix}
 4M \\ 
 k
\end{pmatrix}
\lb \diffz[k + 1]{\mathbb{N}} \rb
\lb \diffz[4M-k]{\mathbb{D}} \rb \\
&= 
\begin{pmatrix}
 4M \\ 
 2M - 1
\end{pmatrix}
\lb \diffz[2M]{ \mathbb{N}} \rb \lb  \diffz[2M + 1]{\mathbb{D}}  \rb
+
\begin{pmatrix}
 4M \\ 
 2M
\end{pmatrix}
\lb \diffz[2M + 1]{\mathbb{N}} \rb \lb \diffz[2M]{\mathbb{D}} \rb,
\end{aligned}
\end{equation}
where the only surviving terms of the above summation are when
$k = 2M-1$, and $k = 2M$. Similarly for the second term in the
numerator in \cref{eq:mfpz_n_d} the surviving terms are obtained when
$k = 2M$ and $k = 2 M + 1$, giving
\begin{equation}
\begin{aligned}
  \label{}
  \diffz[4M]{\ls  \lb  \diffz \mathbb{D} \rb \mathbb{N} \rs }
  &=
\sum_{k = 0}^{4M} 
\begin{pmatrix}
 4M \\ 
 k
\end{pmatrix}
\lb \diffz[k]{\mathbb{N}} \rb
\lb \diffz[4M-k + 1]{\mathbb{D}} \rb \\
&= 
\begin{pmatrix}
 4M \\ 
 2M 
\end{pmatrix}
\lb \diffz[2M]{\mathbb{N}}\rb \lb \diffz[2M + 1]{\mathbb{D}}  \rb
+
\begin{pmatrix}
 4M \\ 
 2M + 1
\end{pmatrix}
\lb \diffz[2M+1]{\mathbb{N}} \rb \lb  \diffz[2M ]{\mathbb{D}} \rb .
\end{aligned}
\end{equation}
Putting it all together gives
\begin{equation}
  \label{eq:mfpz_n_d_lop}
  \ld \diffz \rdfpzz{\nvec}{\novec} \rp_{z = 1} = \ld 
  \frac{
    \lb \diffz[2M + 1]{\mathbb{N}} \rb \lb \diffz[2M]{\mathbb{D}}  \rb
    -
    \lb \diffz[2M]{\mathbb{N}}\rb \lb  \diffz[2M + 1]{\mathbb{D}}  \rb
  }
  {(2M + 1) \ls \diffz[2M]{\mathbb{D}} \rs^{2}} \rp_{z \to 1}.
\end{equation}
Considering the term $\diffz[2M]{\mathbb{N}}$ we find that
\begin{align}
  \label{eq:diff_2m_n}
  \ld \diffz[2M]{\mathbb{N}} \rp_{z \to 1}&= \ld \diffz[2M]{\ls \lb \fpzz{\nvec}{\novec} -1 \rb \mdet{\hmatf} \rs} \rp_{z \to 1}+ \ld \diffz[2M]{\mdet{\hcnmatf[0]}} \rp_{z \to 1}\\
  \label{eq:diff_2m_n_simp}
   &= \ld \diffz[2M]{\mdet{\hcnmatf[0]}} \rp_{z \to 1},
\end{align}
since $\ls \fpzz{\nvec}{\novec} - 1 \rs$ must be differentiated at
least once and $\mdet{\hmatf}$ must be differentiated at least $2M$
times to give a non-zero contribution. From
\cref{eq:hcn0mat_fpz,eq:hcnmat_fpz} we observe that a non-zero
contribution from $\hcnmatf[0]_{i, j}$, $\hcnmatf_{i, j}$ occurs when
one differentiates the difference of first-passage probability,
$\fpzzdif{\nvec}{\vs_j}{\vps_j}$, and the return probability terms,
$\ls 1 - \retzz{\vs_i}\rs$ and $\ls 1 - \retzz{\vps_i}\rs$ at least
once. As
$\fpzsymbol_{\novec}(\vs_i, z = 1) = \fpzsymbol_{\novec}(\vps_i, z =
1) = \fpzsymbol_{\nvec}(\vs_i, z = 1) = \fpzsymbol_{\nvec}(\vps_i, z =
1) = 1$, we have
\begin{equation}
\ld \diffz[2M]{\mathbb{N}} \rp_{z \to 1}  = \ld \diffz[2M]{\hcnmatf[0]} \rp_{z \to 1} =  \ld \diffz[2M]{\hcnmatf} \rp_{z \to 1} = \ld \diffz[2M]{\mathbb{D}} \rp_{z \to 1},
\end{equation}
and we can simplify \cref{eq:mfpz_n_d_lop} to
\begin{equation}
  \label{eq:mfpz_n_d_lop_s}
  \ld \diffz \rdfpzz{\nvec}{\novec} \rp_{z = 1} = \ld 
  \frac{
    \diffz[2M + 1]{\mathbb{N}} 
    -
      \diffz[2M + 1]{\mathbb{D}} 
  }
  {(2M + 1)  \diffz[2M]{\mathbb{D}} } \rp_{z \to 1}.
\end{equation}

At this stage one can compute the derivatives explicitly, for the case
$\diffz[2M]{\hmatf}$ in \cref{eq:det_deriv}. With all $k_i = 2$,
differentiating each column twice and taking the limit $z \to 1$,
gives (after cancelling the $2^M$ term),
\begin{equation}
\label{}
\ld \diffz[2M]{\hmatf} \rp_{z \to 1} = (-1)^{M}(2 M )! \mdet{\hmatdf},
\end{equation}
where
\begin{equation}
  \label{eq:mean_hij_1}
\hmatdf_{i, j} = 
 \rjq[i] \mrett{\vps_i} \mfpttdif{\vs_i}{\vs_j}{\vps_j}
 - \ljq[i] \mrett{\vs_i} \mfpttdif{\vps_i}{\vs_j}{\vps_j}
+ \delta_{i, j} \mrett{\vs_i} \mrett{\vps_i},
\end{equation}
and likewise for the other two matrices we find
\begin{equation}
\label{eq:hnhn0_2mdiff}
\ld \diffz[2M]{\hcnmatf} \rp_{z \to 1} = 
\ld \diffz[2M]{\hcnmatf[0]} \rp_{z \to 1} = 
(-1)^{M}(2 M )! \mdet{\hmatdf -\hsmatdf[2]},
\end{equation}
where
\begin{equation}
  \label{eq:mean_h2ij_1}
\hsmatdf[2]_{i, j} = 
\lb \rjq[i] \mrett{\vps_i}  - \ljq[i] \mrett{\vs_i} \rb \mfpttdif{\nvec}{\vs_j}{\vps_j}.
\end{equation}
Let us now consider the term $\diffz[2 M + 1] \mathbb{N}$. We apply
Leibniz rule and obtain
\begin{equation}
  \label{}
  \diffz[2 M + 1] \mathbb{N}
  = \sum_{\ell = 0}^{2 M + 1}
  \binom{2M + 1}{\ell}
  \ls\bigg.\diffz[\ell]{\lb \fpzz{\nvec}{\novec} - 1\rb} \rs
  \ls \bigg. \diffz[2 M + 1 - \ell]{\mdet{\hmatf}} \rs+
  \diffz[2 M + 1]{\mdet{\hcnmatf[0]}},
\end{equation}
the only surviving term in the summation is when $\ell = 1$ resulting in 
\begin{equation}
  \label{}
  \diffz[2 M + 1] \mathbb{N} = (-1)^{M}\lb 2 M + 1\rb! \mfptt{\nvec}{\novec} \mdet{\hmatdf} + \diffz[2 M + 1]{\mdet{\hcnmatf[0]}}.
\end{equation}
Substituting the previous results into \cref{eq:mfpz_n_d_lop} and
simplifying yields
\begin{equation}
  \label{eq:mfpz_lop_simp_2}
\ld \diffz \rdfpzz{\nvec}{\novec} \rp_{z = 1} = \ld 
    \frac{
{(-1)^{M}\lb 2 M + 1\rb! \mfptt{\nvec}{\novec} \mdet{\hmatdf} + \diffz[2 M + 1]{\mdet{\hcnmatf[0]}}} 
-
\diffz[2 M + 1]{\mdet{\hcnmatf}}
 }
 {(-1)^{M}(2M + 1) !\mdet{\hmatdf -\hsmatdf[2]}}
 \rp_{z \to 1} .
\end{equation}
Consider the term $\diffz[2 M + 1]{\mdet{\hcnmatf[0]}}$ and
$\diffz[2 M + 1]{\mdet{\hcnmatf}}$, using the multinomial expansion of
the derivative of a determinant i.e. \cref{eq:det_deriv}, one can see
that the only surviving terms appear when $M-1$ columns are
differentiated twice, while one column is differentiated three times,
hence,
\begin{align}
  \label{eq:diff_mp1}
\diffz[2 M + 1]{\lc \mdet{\hcnmatf[0]} - \mdet{\hcnmatf} \rc }
&=
\frac{1}{3}\frac{(2 M + 1)!}{2^M}
\sum_{\ell = 1}^{M}
\mdet{\mmat{\widehat{\hmatdf}}^{({\novec}, \ell)}} - 
\mdet{\mmat{\widehat{\hmatdf}}^{({\nvec}, \ell)}},
\end{align}
where
\begin{align}
  \label{}
\mmat{\widehat{\hmatdf}}^{({\novec}, \ell)}_{i, j}
&= 2\lb \delta_{j, \ell}  - 1\rb \lc \hmatdf_{i, j} - \hsmatdf[2]_{i, j} \rc 
+ \delta_{j, \ell}\lc \ld \diffz[3]{\hcnmatf[0]_{i, j}}  \rp_{z = 1} \rc,
\end{align}
\begin{align}
\mmat{\widehat{\hmatdf}}^{({\nvec}, \ell)}_{i, j}
&= 2\lb  \delta_{j, \ell} - 1 \rb \lc \hmatdf_{i, j} - \hsmatdf[2]_{i, j} \rc 
+ \delta_{j, \ell}\lc \ld \diffz[3]{\hcnmatf_{i, j}}  \rp_{z = 1} \rc.
\end{align}
The multilinear property of the determinant allows us to rewrite
\cref{eq:diff_mp1} as
\begin{align}
  \label{eq:diff_mp2}
\diffz[2 M + 1]{\lc \mdet{\hcnmatf[0]} - \mdet{\hcnmatf} \rc }
&=
\frac{1}{3}\frac{(2 M + 1)!}{2^M}
\sum_{\ell = 1}^{M}
\mdet{\mmat{\widehat{\hmatdf}}^{(\ell)}},
\end{align}
where
\begin{align}
  \label{eq:diff_mp3}
\mmat{\widehat{\hmatdf}}^{(\ell)}_{i, j}
&= 2\lb  \delta_{j, \ell} - 1 \rb \lc \hmatdf_{i, j} - \hsmatdf[2]_{i, j} \rc 
+ \ld \delta_{j, \ell}  \diffz[3]{\lc \hcnmatf[0]_{i, j} - \hcnmatf_{i, j} \rc}  \rp_{z = 1} .
\end{align}
Since
\begin{equation}
  \label{eq:hmat_diff}
  \begin{aligned}
{\hcnmatf[0]}_{i, j} - \hcnmatf_{i, j} =
   \fpzzdif{\nvec}{\vs_j}{\vps_j}
  & \lc \rjq[i]\ls 1 - \retzz{\vps_i} \rs\fpzzdif{\vs_i}{\novec}{\nvec} \rd \\
-  &\ljq[i] \ld \ls 1 -\retzz{\vs_i} \rs \fpzzdif{\vps_i}{\novec}{\nvec} \rc,
\end{aligned}                                       
\end{equation}
each of the first-passage and return probabilities in
\cref{eq:hmat_diff} must be
differentiated at-least once to give a non-zero contribution, hence,
\begin{align}
\ld \diffz[3]{\lc \hcnmatf[0]_{i, j} - \hcnmatf_{i, j} \rc} \rp_{z = 1} &=
   6 {\mfpttdif{\nvec}{\vs_j}{\vps_j} \lc
 \rjq[i]  \mrett{\vps_i} \mfpttdif{\vs_i}{\novec}{\nvec}
- \ljq[i]\mrett{\vs_i}  \mfpttdif{\vps_i}{\novec}{\nvec} \rc},
\label{eq:diff_mp_diff3_sol}
\end{align}
where the factor 6 comes from repeated application of the product rule.
To carry out the summation in \cref{eq:diff_mp2} we employ the
following property of determinants. Given two matrices, $\mmat{A}$ and
$\mmat{B}$ of size $M \times M$, where
$\mmat{B} = \mvec{a}\mvec{b^{\trans}}$ and where $\mvec{a} $ and
$\mvec{b}$ are two column vectors, the following relation
\begin{equation}
  \label{eq:mat_sum_identity}
\sum_{\ell = 1}^{M} \mdet{\mmat{A}^{(\ell)}} = \mdet{\mmat{A}} - \mdet{\mmat{A} - \mmat{B}} 
\end{equation}
holds when $\mmat{A}^{(\ell)}$ is the same as $\mmat{A}$, but with the
$\ell^{\thh}$ column replaced by the $\ell^{\thh}$ column of
$\mmat{B}$. Comparing \cref{eq:diff_mp2,eq:diff_mp_diff3_sol} with
\cref{eq:mat_sum_identity}, we observe that
$\mmat{A} \to \hmatdf - \hsmatdf[2]$,
\begin{equation}
  \begin{aligned}
    &\mvec{a} \to \ls \rjq[1] \mrett{\vps_1}\mfpttdif{\vs_1}{\novec}{\nvec} - \ljq[1] \mrett{\vs_1}\mfpttdif{\vs_1}{\novec}{\nvec}, \rd \\
    &\phantom{\mvec{a} \to \ls \rd} \ld \cdots, \rjq[M] \mrett{\vps_M}\mfpttdif{\vs_M}{\novec}{\nvec} - \ljq[M] \mrett{\vs_M}\mfpttdif{\vs_M}{\novec}{\nvec}\rs^{\trans},
  \end{aligned}
\end{equation}
and
\begin{equation}
  \mvec{b} \to 6\ls \mfpttdif{\nvec}{\vs_1}{\vps_1}, \cdots,
  \mfpttdif{\nvec}{\vs_M}{\vps_M} \rs^{\trans},
\end{equation}
and carrying out the summation we obtain
\begin{align}
  \label{eq:diff_mp4}
\frac{1}{3}\frac{(2 M + 1)!}{2^M}
\sum_{\ell = 1}^{M}
\mdet{\mmat{\widehat{\hmatdf}}^{(\ell)}} &=
\frac{1}{3}\frac{(2 M + 1)!}{2^M} \lc \bigg.
  \mdet{-2 \lb \hmatdf - \hsmatdf[2] \rb}
  - 
  \mdet{-2 \lb \hmatdf - \hsmatdf[2] \rb - 6 \hsmatdf[1]} \rc,
\end{align}
with
\begin{align}
  \label{eq:mean_h1ij}
  \hsmatdf[1]_{i, j} &= \lb 
                       \rjq[i] \mrett{\vps_i} \mfpttdif{\vs_i}{\novec}{\nvec} -  
                       \ljq[i] \mrett{\vs_i} \mfpttdif{\vps_i}{\novec}{\nvec}  \rb
                       \mfpttdif{\nvec}{\vs_j}{\vps_j}.
\end{align}
From \cref{eq:diff_mp4}, we can factor out (-2) from the determinants
using the property $\mdet{\alpha \mmat{A}} = \alpha^{M} \mdet{\mmat{A}}$ with
$\mmat{A}$ an $M \times M $ determinant to yield
\begin{align}
  \label{eq:diff_mp5}
\frac{1}{3}\frac{(2 M + 1)!}{2^M}
\sum_{\ell = 1}^{M}
\mdet{\mmat{\widehat{\hmatdf}}^{(\ell)}} =
\frac{(-1)^{M}(2 M + 1)!}{3} \lc \bigg. \mdet{\lb \hmatdf - \hsmatdf[2] \rb} - \mdet{ \lb \hmatdf - \hsmatdf[2] \rb + 3 \hsmatdf[1]} \rc.
\end{align}
The factor $1/3$ can be taken into the determinants using the property
given in \cref{eq:det_prop}. For the first determinant on the right
hand side (RHS) of \cref{eq:diff_mp5} we can equate
$\mmat{A} \to \hmatdf$, from \cref{eq:mean_h2ij_1} we can equate
\begin{equation}
  \label{eq:a_mean_1}
\mvec{a} \to \ls \rjq[1] \mrett{\vps_1}  - \ljq[1] \mrett{\vs_1}, \cdots, \rjq[M] \mrett{\vps_M}  - \ljq[1] \mrett{\vs_M}   \rs^{\trans}.
\end{equation}
and
\begin{equation}
  \label{eq:b_mean_1}
\mvec{b} \to \ls \mfpttdif{\nvec}{\vs_1}{\vps_1}, \cdots, \mfpttdif{\nvec}{\vs_M}{\vps_M} \rs^{\trans},
\end{equation}
to derive the relation
\begin{equation}
  \label{eq:det_expand_1}
\frac{1}{3}\mdet{\hmatdf - \hsmatdf[2]} = \mdet{\hmatdf - \sfrac{1}{3} \hsmatdf[2]} - \frac{2}{3}\mdet{\hmatdf}.
\end{equation}
Similarly, for the second determinant on the RHS of
\cref{eq:diff_mp5}, we can equate $\mmat{A} \to \hmatf$, and using
\cref{eq:mean_h1ij} we identify
\begin{equation}
  \label{eq:a_mean_2}
  \begin{aligned}
  &\mvec{a} \to \ls \rjq[1] \mrett{\vps_1} \lb 1 - 3\mfpttdif{\vs_1}{\novec}{\nvec} \rb
    - \ljq[1] \mrett{\vs_1}\lb 1 - 3\mfpttdif{\vps_1}{\novec}{\nvec} \rb, \rd \\
    &\phantom{\mvec{a} \to \ls \rd} \ld \cdots,
  \rjq[M] \mrett{\vps_M} \lb 1 - 3\mfpttdif{\vs_M}{\novec}{\nvec} \rb - \ljq[1] \mrett{\vs_M}\lb 1 - 3\mfpttdif{\vps_M}{\novec}{\nvec} \rb   \rs^{\trans},
  \end {aligned}
\end{equation}
and
\begin{equation}
  \label{eq:b_mean_2}
\mvec{b} \to \ls \mfpttdif{\nvec}{\vs_1}{\vps_1}, \cdots, \mfpttdif{\nvec}{\vs_M}{\vps_M} \rs^{\trans},
\end{equation}
to obtain the relation
\begin{equation}
  \label{eq:det_expand_2}
\frac{1}{3}\mdet{\hmatdf - \hsmatdf[2] + 3 \hsmatdf[1]} = \mdet{\hmatdf - \sfrac{1}{3} \hsmatdf[2] + \hsmatdf[1]} - \frac{2}{3}\mdet{\hmatdf}.
\end{equation}
Using the relations \cref{eq:det_expand_1,eq:det_expand_2} to simplify
\cref{eq:diff_mp5} yields
\begin{align}
  \label{eq:diff_mp6}
\frac{1}{3}\frac{(2 M + 1)!}{2^M}
\sum_{\ell = 1}^{M}
\mdet{\mmat{\widehat{\hmatdf}}^{(\ell)}} 
  &=(-1)^M(2M + 1)!\lc \mdet{\hmatdf - \sfrac{1}{3}\hsmatdf[2]} - \mdet{ \hmatdf - \sfrac{1}{3}\hsmatdf[2] + \hsmatdf[1]} \rc.
\end{align}
Lastly, employing the property 
\begin{equation}
  \mdet{\mmat{A} - \mvec{a} \mvec{b}^{ \trans }} - \mdet{\mmat{A} - \mvec{c} \mvec{b}^{\trans}}  = 
  \mdet{\mmat{A} - \lb \mvec{a} -  \mvec{c} \rb  \mvec{b}^{\trans}} - \mdet{\mmat{A}}
  \label{eq:mat_sec_prop}
\end{equation}
with $\mmat{A} \to \hmatdf$, $3 \mvec{a}$ with \cref{eq:a_mean_1},
$3\mvec{b}$ with \cref{eq:b_mean_1} and $3\mvec{c}$  with
\cref{eq:a_mean_2}, we can simplify \cref{eq:diff_mp6} to
\begin{align}
  \label{eq:diff_mp7}
\frac{1}{3}\frac{(2 M + 1)!}{2^M}
\sum_{\ell = 1}^{M}
\mdet{\mmat{\widehat{\hmatdf}}^{(\ell)}} 
  =(-1)^M(2M + 1)!\lc \mdet{\hmatdf - \hsmatdf[1]} - \mdet{ \hmatdf} \rc.
\end{align}
Putting it all together gives the final
mean first-passage time with $M$ paired defects
\begin{equation}
  \label{eq:rdmfpt_o}
\begin{split}
\rdmfpt_{\novec \to \nvec}
= \frac
{\lb \mfptt{\nvec}{\novec} - 1 \rb \mdet{\hmatdf} + \mdet{\hmatdf - \hsmatdf[1]}}
{\mdet{\hmatdf - \hsmatdf[2]}} 
\end{split}
\end{equation}
If we divide through all the terms by $\prod_{(\vs, \vps) \in
  S}\mrett{\vs}\mrett{\vps}$, we obtain
\begin{equation}
  \label{eq:rdmfpt}
\begin{split}
\rdmfpt_{\novec \to \nvec}
= \frac {\mfptt{\nvec}{\novec} \mdet{\hmatd - 1/\mfptt{\nvec}{\novec}\hsmatd[1]}}
{\mdet{\hmatd - \hsmatd[2]}},
\end{split}
\end{equation}
where
\begin{align}
  \label{eq:mean_hij_final}
\hmatd_{i, j} &= \frac{\rjq[i]}{\mrett{\vs_i}} \mfpttdif{\vs_i}{\vs_j}{\vps_j} -
                \frac{\ljq[i]}{\mrett{\vps_i}} \mfpttdif{\vps_i}{\vs_j}{\vps_j}
+ \delta_{i, j},\\
  \label{eq:mean_h1ij_final}
\hsmatd[1]_{i, j} &= \lb 
\frac{\rjq[i]}{\mrett{\vs_i}}  \mfpttdif{\vs_i}{\novec}{\nvec} -  
\frac{\ljq[i]}{\mrett{\vps_i}} \mfpttdif{\vps_i}{\novec}{\nvec}  \rb
\mfpttdif{\nvec}{\vs_j}{\vps_j},\\
  \label{eq:mean_h2ij_final}
  \hsmatd[2]_{i, j} &= \lb \frac{\rjq[i]}{\mrett{\vs_i}}   -
                      \frac{\ljq[i]}{\mrett{\vps_i}} 
                      \rb \mfpttdif{\nvec}{\vs_j}{\vps_j} .
\end{align}
These expressions, namely
\cref{eq:mean_hij_final,eq:mean_h1ij_final,eq:mean_h2ij_final}, have
been repeated, respectively, in
\cref{mt-eq:mean_hij_final,mt-eq:mean_h1ij_final,mt-eq:mean_h2ij_final}
of the main text. If the homogeneous propagator is diffusive with no
bias and if the heterogeneity parameters are symmetric, i.e.
$\rjq = \ljq$, $\hsmatd[2] = 0$ and \cref{eq:rdmfpt} can be simplified
further to
\begin{equation}
  \label{eq:rdmfpt_sym}
  \rdmfptt{\nvec}{\novec}
  =
  \mfptt{\nvec}{\novec} - 1 
  + 
  \frac
  {
    \mdet{\hmatd - \hsmatd[1]}
  }
  {
    \mdet{\hmatd}
  }.
\end{equation}






\subsection{Mean first-passage in the presence of sticky or slippery sites}
To build the first-passage probability with sticky and slippery
heterogeneities (see \cref{sec:sticky_slippery_prop} and
\cref{fig:sticky_defect_schematic}) we use the propagator given in
\cref{mt-eq:rdprop_sol_mt}, where the matrices $\hmat$ and
$\hcnmat[0]$ are given, respectively, by
\cref{eq:hmat_sticky,eq:hcn0mat_sticky}. As the procedure is
similar to the one used to derive the general MFPT given by
\cref{eq:rdmfpt} (and also \cref{mt-eq:rdmfpt_mt} in the main text),
we outline only the key steps.

Starting from
\begin{equation}
  \label{eq:rdfpz_props_sticky}
  \rdfpz_{\novec}(\nvec, z) =
   \frac{\rdpropz_{\novec}(\nvec, z)}{\rdpropz_{\nvec}(\nvec, z)} = 
\frac {\lb \propz_{\novec}(\nvec, z) - 1 \rb \mdet{\hmat} + \mdet{\hmat - \hsmat[0]}}
{\lb \propz_{\nvec}(\nvec, z) - 1 \rb \mdet{\hmat} + \mdet{\hmat - \hsmat}},
\end{equation}
where we have called $\hcnmat[0] = \hmat - \hsmat[0]$, with $\hmat$
and $\hcnmat[0]$ given, respectively, by
\cref{eq:hmat_sticky,eq:hcn0mat_sticky}. Since matrix
$\hsmatfull{\mvec{n}}{\mvec{m}} = \mvec{a}\mvec{b}^{\trans}$, where
$\mvec{a}$ and $\mvec{b}$ are column vectors with elements
$\mvec{a}_i = \propzz{\ws_i}{\mvec{m}}$ and
$\mvec{b}_i = \sypropzz{\nvec}{\ws_i}$ with $\sypropzz{\nvec}{\ws}$
given by \cref{eq:omega_ap}.  Dividing both the numerator and
denominator of \cref{eq:rdfpz_props_sticky} by
$\propzz{\nvec}{\novec} \prod_{i = 1}^{M} \propzz{\ws_i}{\ws_i}$ we
find
\begin{equation}
  \label{eq:rdfpz_fpz_sitcky}
 \rdfpz_{\novec}(\nvec, z) =
\frac {\lb \fpzz{\nvec}{\novec}   - 1\rb \mdet{\hmatf} + \mdet{\hcnmatf[0]}}
{\mdet{\hcnmatf}}.
\end{equation}
where we rewrite the elements of the matrices in terms of
first-passage and first return probabilities giving the definitions
\begin{align}
  \label{eq:hmat_fpz_sticky}
  \hmatf_{i, j} &=
  \sypropzz{\ws_i}{\ws_j} - z^{-1}\delta_{i, j}\ls 1 -  \retzz{\ws_i} \rs, \\
  \label{eq:hcnmat_fpz_sticky}
   \hcnmatf_{i, j} &= \hmatf_{i, j} - \fpzz{\ws_i}{\nvec}\sypropzz{\nvec}{\ws_j}, \\
  \label{eq:hcn0mat_fpz_sticky}
   \hcnmatf[0]_{i, j} &= \hmatf_{i, j} - \fpzz{\ws_i}{\novec}\sypropzz{\nvec}{\ws_j},
\end{align}
with 
\begin{equation}
  \sypropzz{\nvec}{\ws_j} = 
\sum_{s = 1}^{d} 
\rjq[][{\wsr[j]}][{\ws_j}]
\fpzzdif{\nvec}{\ws_j}{\wsr[j]} + 
\ljq[][{\wsl[j]}][{\ws_j}]
\fpzzdif{\nvec}{\ws_j}{\wsl[j]}.
\label{eq:omega_fpz}
\end{equation}
Note that \cref{eq:hmat_fpz,eq:hcn0mat_fpz,eq:hcnmat_fpz} are now
different from
\cref{eq:hmat_fpz_sticky,eq:hcn0mat_fpz_sticky,eq:hcnmat_fpz_sticky}.

The derivative with respect to $z$ yields \cref{eq:mfpz_n_d}, however,
to evaluate the limit $z \to 1$, one must employ de L'Hôpital's rule
$2M$ times on \cref{eq:mfpz_n_d}, where previously the rule was used
$4M$ times. This is because the elements of the matrices given by
\cref{eq:hmat_fpz_sticky,eq:hcnmat_fpz_sticky,eq:hcn0mat_fpz_sticky}
need only be differentiated once to give non-zero contributions, i.e
the determinants, $\mdet{\hmat}, \mdet{\hcnmat}$ and
$\mdet{\hcnmat[0]}$ must be differentiated $M$ times. Following
through one finds, instead \cref{eq:mfpz_n_d_lop_s},
\begin{equation}
  \label{eq:mfpz_n_d_lop_s}
  \ld \diffz \rdfpzz{\nvec}{\novec} \rp_{z = 1} = \ld 
  \frac{
    \diffz[M+1]{\mathbb{N}} 
    -
      \diffz[M+1]{\mathbb{D}} 
  }
  {(M+1)  \diffz[M]{\mathbb{D}} } \rp_{z \to 1}.
\end{equation}
After computing the derivatives explicitly one obtains
\cref{eq:rdmfpt} but this time
\begin{align}
  \label{eq:mean_hij_1_sticky}
\hmatd_{i, j} &= \frac{1}{\mrett{\ws_i}} \symfptt{\ws_i}{\ws_j}
+ \delta_{i, j},\\
  \label{eq:mean_h1ij_1_sticky}
\hsmatd[1]_{i, j} &= \frac{1}{\mrett{\ws_i}}\mfpttdif{\ws_i}{\novec}{\nvec} \symfptt{\nvec}{\ws_j}, \\
  \label{eq:mean_h2ij_1_sticky}
\hsmatd[2]_{i, j} &= \frac{1}{\mrett{\ws_i}} \symfptt{\nvec}{\ws_j},
\end{align}
and with 
\begin{equation}
  \label{eq:omega_mfpt}
  \begin{aligned}
  \symfptt{\nvec}{\ws_j} = 
\sum_{s = 1}^{d} 
\rjq[][{\wsr[j]}][{\ws_j}]
\mfpttdif{\nvec}{\ws_j}{\wsr[j]} + \ljq[][{\wsl[j]}][{\ws_j}]
\mfpttdif{\nvec}{\ws_j}{\wsl[j]}.
  \end{aligned}
\end{equation}

\subsection{Mean first-return time}
\label{sec:mret_deriv}
Through the renewal equation we also have the return probability
relation
\begin{equation}
  \label{eq:rdretz_props}
  \begin{aligned}
  \rdretzz{\nvec} &= 
   1 - \frac{1}{\rdpropz_{\nvec}(\nvec, z)}= 
\frac{\lb \propzz{\nvec}{\nvec} - 2 \rb \mdet{\hmat} + \mdet{\hcnmat}}
{\lb \propzz{\nvec}{\nvec} - 1 \rb \mdet{\hmat} + \mdet{\hcnmat}}.
  \end{aligned}
\end{equation}
Dividing both the numerator and denominator of \cref{eq:rdretz_props}
by $\propz_{\nvec}(\nvec, z) \prod_{i = 1}^{M} \propzz{\ws_i}{\ws_i}$
and simplifying gives
\begin{align}
  \label{eq:rdretz_fpz}
  \rdretzz{\nvec} &= 
   \frac {\lb \retzz{\nvec} - 1\rb \mdet{\hmatf}
     + \mdet{\hcnmatf}}
   {\mdet{\hcnmatf}},
\end{align}
where $\hmatf, \hcnmatf[0]$ and $\hcnmatf$ are given by,
\cref{eq:hmat_fpz} , \cref{eq:hcn0mat_fpz} and \cref{eq:hcnmat_fpz} in
the case of the general paired defect. Whereas for the sticky-slippery
defects one would use \cref{eq:hmat_fpz_sticky},
\cref{eq:hcn0mat_fpz_sticky} and \cref{eq:hcnmat_fpz_sticky}. In
either case the \cref{eq:rdretz_fpz} is structurally identical to
\cref{eq:rdfpz_fpz} and one employs the same procedure as the one
employed to derive the MFPT to obtain (also given in
\cref{mt-eq:rdmret_mt} in the main text)
\begin{equation}
  \label{eq:rdmret_sup}
\mathfrak{R}_{\nvec}
=
\frac
{
\mrett{\nvec}
\mdet{\hmatd}
}
{
\mdet{\hmatd - \hsmatd[2]}
}
\end{equation}
where $\hmatd$, $\hsmatd[2]$ are given, respectively, by
\cref{mt-eq:mean_hij_final} and \cref{mt-eq:mean_h2ij_final} in the
case of paired defects, while for sticky-slippery defects one would
use the definitions in \cref{eq:mean_hij_1_sticky} and
\cref{eq:mean_h2ij_1_sticky}.

\section{Steady State Probability}
\label{sec:steady_state}
Using the final value theorem, the steady state probability is given
by 
\begin{equation}
\begin{aligned}
  \label{}
  &\ld (1 - z)
  \rdpropzz{\nvec}{\novec} \rp_{z \to 1} =  \ld (1 - z) \propzz{\nvec}{\novec} \rp_{z \to 1} \\
  & - \ld \frac{(1 - z)\mdet{\hmat} -(1 - z) \mdet{\hmat - \hsmat[0]} }{\mdet{  \hmat}} \rp_{z \to 1},
\end{aligned}
\end{equation}
taking the limit requires a similar procedure as the one given for the
derivation of the MFPT see \cref{sec:mfpt_deriv}, and after some
algebra one finds
\begin{equation}
  \label{eq:steady_state_sup}
  \ld (1 - z)
\rdpropzz{\nvec}{\novec} \rp_{z \to 1}  = 
\frac{1}{\mrett{\nvec}}
 \frac{ \mdet{\hmatd - \hsmatd[2]}}{\mdet{\hmatd}}
\end{equation}
which (as expected) is the reciprocal of the MRT and where the
elements of $\hmatd$, $\hsmatd[1]$ and $\hsmatd[2]$ are defined,
respectively, in \cref{eq:mean_hij_final}, \cref{eq:mean_h1ij_final}
and \cref{eq:mean_h2ij_final} for the general case, and
\cref{eq:mean_hij_1_sticky}, \cref{eq:mean_h1ij_1_sticky} and
\cref{eq:mean_h2ij_1_sticky} when the heterogeneities are only sticky
and slippery.

\section{First-Passage Statistics In One Dimensional Domains}
\label{sec:expression_1d_suppl}
Here we display some explicit miscellaneous expressions for 1D systems
that we have omitted from the main text.
\subsection{Mean first-passage time in periodic domains}

In a periodic domain when $n \leq u < n_0$ or $n > u \geq n_0$ the
MFPT can be shown to be equal to
\begin{equation}
\rdmfptt{n}{n_0} = \mfptt{n}{n_0} +
\frac{\jqsymbol \lb n - n_0  - \mathrm{sign}(n - n_0)\rb \ls N\, \mathrm{sign}(n - n_0) + 1 - 2 (n - u) \rs}{p \ls N \frac{p}{2} + \jqsymbol (1 - N) \rs},
\end{equation}
while when both $n$ and $n_0$ are to the right or left of the barrier
ones finds
\begin{equation}
\rdmfptt{n}{n_0} = \mfptt{n}{n_0} +  
\frac{\jqsymbol \lb n - n_0 \rb \ls N\, \mathrm{sign}(u - n) + 1 - 2 (n - u) \rs}{p \ls N \frac{p}{2} + \jqsymbol (1 - N) \rs},
\end{equation}
where $\mathrm{sign}(m) = 1$ for $m \geq 0$ and
$\mathrm{sign}(m) = -1$ for $m < 0$ is the discrete signum function.



\subsection{Mean return time in 1D}
\label{sec:mret_limit}
Using \cref{mt-eq:rdmret_mt} from the main text one can show that the
mean return time in 1D with a single barrier at $u$ with periodic
boundary conditions is given by
\begin{equation}
  \label{eq:mret_1d_peri}
  \begin{aligned}
    \rdmrett{n} =  \lc
    \begin{array}{ll}
      \frac{\frac{q}{2} \ls N^2 - \lb N^2 - N\rb\lb\jqsymbol_{u} + \jqsymbol_{\varv} \rb \rs}
      {N \lb \ffrac{q}{2} - \jqsymbol_{\varv} \rb + \jqsymbol_{u}\lb n - u \rb - \jqsymbol_{v} \lb n - u - 1 \rb}, & n \leq u, \\
      & \\
      \frac{\frac{q}{2} \ls N^2 - \lb N^2 - N\rb\lb\jqsymbol_{u} + \jqsymbol_{\varv} \rb \rs}
      {N \lb \ffrac{q}{2} - \jqsymbol_{u} \rb + \jqsymbol_{u}\lb n - u \rb - \jqsymbol_{v} \lb n - u - 1 \rb}, & n \geq u + 1.
      \end{array} \rd
  \end{aligned}
\end{equation}

When the barrier is impenetrable in both directions,
i.e. $\rjq[][u+1][u], \ljq[][u][u+1] \to \frac{q}{2}$, the mean return
time with periodic boundary condition remains $\rdmrett{n} = N$ as we
have transformed the periodic boundary condition to a reflecting
one. However, To recover the analogue with reflecting boundary condition
requires careful consideration of each of the terms in
\begin{equation}
  \label{eq:mret_1d_ref_suppl}
\rdmrett{n}^{(r)} = 
\lc
\begin{array}{ll}
N\ls \frac{\ffrac{q}{2}-\rjq[][u+1][u]}{\ffrac{q}{2} - \ljq[][u][u+1] } \rs
- u \ls \frac{\ljq[][u][u+1] -\rjq[][u+1][u]}{\ffrac{q}{2}- \ljq[][u][u+1]} \rs & n \leq  u  \\
& \\
N - u \ls \frac{\rjq[][u][u+1] - \ljq[][u+1][u]}{\ffrac{q}{2}-\rjq[][u+1][u] } \rs & n \geq u + 1  
\end{array} 
\rd.
\end{equation}
When $n \leq u$, by expanding the first term in \cref{eq:mret_1d_ref_suppl}
it is obvious to see that it gives no contribution to
$\rdmrett{n}^{(r)}$
\begin{equation}
  \label{}
  \begin{aligned}
N\lim_{\substack{\rjq[][u+1][u] \to \frac{q}{2} \\ \ljq[][u][u+1] \to \frac{q}{2}}} \lb \frac{\frac{q}{2} - \rjq[][u+1][u]}{\frac{q}{2} - \rjq[][u][u+1]} \rb
&= \frac{Nq}{2} \lim_{\ljq[][u][u+1] \to \frac{q}{2}}  \lb \frac{1}{\frac{q}{2} - \ljq[][u][u+1]} \rb -\frac{Nq}{2}   \lim_{\ljq[][u][u+1] \to \frac{q}{2}} \lb\frac{1}{\frac{q}{2} - \rjq[][u][u+1]} \rb 
= 0,
  \end{aligned}
\end{equation}
while the second term gives
\begin{equation}
  \label{eq:mret_1d_lim_term_2}
  \begin{aligned}
u\lim_{\substack{\rjq[][u+1][u] \to \frac{q}{2} \\ \ljq[][u][u+1] \to \frac{q}{2}}} \lb \frac{\ljq[][u][u+1] - \rjq[][u+1][u]}{\frac{q}{2} - \rjq[][u][u+1]} \rb
&= \frac{uq}{2}\lim_{\ljq[][u][u+1] \to \frac{q}{2}} \lb\frac{1}{\frac{q}{2} - \rjq[][u][u+1]} \rb  - u \lim_{\ljq[][u][u+1] \to \frac{q}{2}}  \lb \frac{\rjq[][u][u+1]}{\frac{q}{2} - \ljq[][u][u+1]} \rb  
= u.
  \end{aligned},
\end{equation}
similar argument can be made for the case when $n > u$ to give
$\rdmrett{n}^{(r)} = N - u$.

\subsection{Mean exit time}
In 1D, one find simple expressions for the mean exit times, using
known expressions for the defect free exit time
$\mexitt{n_0} = (N - n_0)(n_0 - 1) / q$ \cite{giuggioli2020}, and the
overall survival probability of the 1D diffusive propagator with absorbing
boundaries, given by
\begin{equation}
  \label{}
\lim_{z \to 1}\propzz{n}{n_0} = \frac{1}{q}\frac{2 (N - n_>)(n_< - 1)}{N - 1},
\end{equation}
where $n_> =  \frac{1}{2} \ls \lp n - n_0 \rp  + (n + n_0) \rs $ and
$n_< =  \frac{1}{2} \ls \lp n - n_0 \rp  - (n + n_0) \rs $. When $n_0
\geq u + 1$, we find the simple relatively relation
\begin{equation}
  \label{}
  \rdmexitt{\novec} = \mexitt{\novec}  
+ \frac{ (2u - N) (N - n_0) \ls u (\jqsymbol_u -\jqsymbol_{\varv}) - \jqsymbol_{u} \rs}
{q \ls N(\frac{q}{2} -  \jqsymbol_{\varv}) +  \jqsymbol_u + \jqsymbol_{\varv}  - u ( \jqsymbol_{u} - \jqsymbol_{\varv} ) - \frac{q}{2}\rs},
\end{equation}
whereas $n_0 \leq u $ yields
\begin{equation}
  \label{}
  \rdmexitt{\novec} = \mexitt{\novec}  
+ \frac{ (2u - N) (1 - n_0) \ls (u - N) (\jqsymbol_u -\jqsymbol_{\varv}) - \jqsymbol_{\varv} \rs }
{q \ls N(\frac{q}{2} -  \jqsymbol_{\varv}) +  \jqsymbol_u + \jqsymbol_{\varv}  - u ( \jqsymbol_{u} - \jqsymbol_{\varv} ) - \frac{q}{2}\rs}.
\end{equation}

\section{Efficient Evaluation Of The Propagator In Finite Domains: The
Block Matrix Construction}
\label{sec:block_matrix_const}
When the number of paired defects is sufficiently small, e.g.
$M \lessapprox 10$ it is convenient to compute the elements of the
matrices in \cref{mt-eq:rdprop_sol_mt,mt-eq:rdmfpt_mt} (of the main
text) directly using, respectively, the homogeneous propagators and
mean first-passage times. Whereas, for larger values of $M$ it is more
efficient to evaluate the heterogeneous propagator using a block
matrix construction containing eigenvectors and eigenvalues of the
transition matrix. In what follows we describe the procedure for a 1D
system while the extension to higher dimension will be addressed in
the following subsection. We define the matrices containing the right
eigenvectors as
\begin{align}
  \label{eq:R}
\mmat{R}_{i, k} = g^{(\gamma, r)}(\vsc_i, k)
\quad\mathrm{and}\quad 
\mmat{R}'_{i, k} = g^{(\gamma, r)}(\vpsc_i, k),
\end{align}
where $g^{(\gamma, r)}(\vsc_i, k)$ and $g^{(\gamma, r)}(\vpsc_i, k)$
are, respectively the $\vsc_i^{\mathrm{th}}$ and
$\vpsc_i^{\mathrm{th}}$ component of the $k^{\mathrm{th}}$ right
eigenvector given by \cref{eq:gr}, with the type of boundary
condition described by $\gamma$. Similarly for the matrices containing
the left eigenvectors we define
\begin{align}
  \label{eq:L}
\mmat{L}_{k, i} = g^{(\gamma, \ell)}(\vsc_i, k) 
\quad\mathrm{and}\quad 
\mmat{L}'_{k, i} = g^{(\gamma, \ell)}(\vpsc_i, k),
\end{align}
where $g^{(\gamma, \ell)}(\vsc_i, k)$ and
$g^{(\gamma, \ell)}(\vpsc_i, k)$ are, respectively, the
$\vsc_i^{\mathrm{th}}$ and $\vpsc_i^{\mathrm{th}}$ component of the
$k^{\mathrm{th}}$ right eigenvector given by \cref{eq:gr}, while for
the matrices of eigenvalues we define a diagonal matrix $\mmat{K}$
with elements $\mmat{K}_{k, k} = s^{(\gamma)}(k)$ where $s^{(\gamma)}(k)$ is the
$k^{\mathrm{th}}$ eigenvalue of the homogeneous system given by
e.g. Eq.(4) of Ref.~\cite{giuggioli2020} or Eq.~(22) of
Ref.~\cite{sarvaharmangiuggioli2020}. For the dependence on the the
occupation site $n$ and the initial site $n_0$ we define the row
vector $\mvec{r}_n$ and the column $\mvec{\ell}_{n_0}$ containing,
respectively, the $n^{\mathrm{th}}$ and ${n_0}^{\mathrm{th}}$
components of the right and left eigenvectors. Finally we define the
diagonal matrices
\begin{align}
  \label{eq:Lambda}
\mmat{\Lambda}_{i, i} = \rjq[i][\vsc][\vpsc]
\quad\mathrm{and}\quad 
\mmat{\Lambda}'_{i, i} = \ljq[i][\vpsc][\vsc],
\end{align}
which contain all the heterogeneous parameter values. The size of
these matricies depend on the number of heterogeneties, $M$, the size
of the domain $N$ or both: the matricies $\mmat{R}$ and $\mmat{R}'$
are of size $M \times N $; the matricies $\mmat{L}$ and $\mmat{L}'$
are of size $N \times M $; the matricies $\mmat{\Lambda}$ and
$\mmat{\Lambda}'$ are of size $M \times M$; and lastly $\mmat{K}$ is
of size $N \times N$. From these definitions it follows that
\begin{equation}
  \label{}
  \hmat = \mmat{Y} - z^{-1}\mmat{I}
  \quad\mathrm{and}\quad 
  \hsmatfull{n}{n_0} = \mvec{o} \mvec{s}
\end{equation}
where
\begin{align}
  \label{}
  \mmat{Y} &=
\ls \mmat{\Lambda} \mmat{R} - \mmat{\Lambda}' \mmat{R}' \rs \ls \mmat{I} - z \mmat{K} \rs^{-1} \ls \mmat{L} - \mmat{L}' \rs \\
  \mvec{o} &= 
  \ls \mmat{\Lambda} \mmat{R} - \mmat{\Lambda}' \mmat{R}' \rs \ls \mmat{I} - z \mmat{K} \rs^{-1} \mvec{\ell}_0 \\
  \mvec{s} &= \mvec{r}_n \ls \mmat{I} - z \mmat{K} \rs^{-1} \ls \mmat{L} - \mmat{L}' \rs,
\end{align}
By defining the block matrices
\begin{equation} 
\begin{matrix}
\mmat{X}_L = \begin{pmatrix} \mmat{L} - \mmat{L}' & \mvec{l}_0 \end{pmatrix} & \text{and} &
\mmat{X}_R = \begin{pmatrix} \mmat{u}\mmat{R} - \mmat{u}'\mmat{R}' \\ \mvec{r}_n \end{pmatrix}
\end{matrix}
\end{equation} 
we find
\begin{equation}
  \mmat{X}_R \ls \mmat{I} - z \mmat{K} \rs^{-1} \mmat{X}_L =
  \begin{pmatrix}
   \mmat{Y} & \mvec{o} \\
   \mvec{s}^{\trans} & \propzz{n}{n_0}
  \end{pmatrix}.
\end{equation}
A similar approach can be used for the matrices involved MFPT given by \cref{eq:rdmfpt}.

\subsection{The block matrix construction in higher dimensions}
In higher dimension the transition dynamics of a lattice random walk
are described by a tensor and extending the block matrix construction
hinges on `flattening' the vector coordinates to a scalar. There are
many methods one can employ to achieve this and we outline a suitable
one below.  Given the site $\nvec = (n_1, \cdots, n_d)$ in a
$d$-dimensional lattice of size $\mvec{N} = N_1, \cdots N_d$, we
define
\begin{equation}
  \label{}
\hat{n} = 1 + \sum_{i = 1}^{d} \lb \prod_{j = 1}^{i - 1} N_{j} \rb \lb n_i - 1\rb \quad \text{and} \quad
\hat{k} = 1 + \sum_{i = 1}^{d} \lb \prod_{j = 1}^{i - 1} N_{j} \rb \lb k_i  -1\rb,
\end{equation}
where $\hat{n}$ represents the `flattened' site while $\hat{k}$ is the
`flattened' eigen-index. Using these indices we define
\begin{align}
  \label{eq:Rh}
\mmat{R}_{i, \hat{k}} = \prod_{j = 1}^{d} g_{r}^{(\gamma_j)}(\vsc_{i_j}, k_{_j})
\quad\mathrm{and}\quad 
\mmat{R}'_{i, \hat{k}} = \prod_{j = 1}^{d} g_{r}^{(\gamma_j)}(\vpsc_{i_j}, k_{_j}),
\end{align}
where the products ares over $j^{\thh}$ component of the sites,
e.g. $\vs_i = (\vsc_{i_1}, \cdots, \vsc_{i_d})$. Similarly, the
matricies containing the left eigenvectors are defined as 
\begin{align}
  \label{eq:Rh}
\mmat{L}_{\hat{k}, i} = \prod_{j = 1}^{d} g_{\ell}^{(\gamma_j)}(\vsc_{i_j}, k_{_j})
\quad\mathrm{and}\quad 
\mmat{L}'_{\hat{k}, i} = \prod_{j = 1}^{d} g_{\ell}^{(\gamma_j)}(\vpsc_{i_j}, k_{_j}),
\end{align}
leaving the matrix of eigenvalues defined as
$\mmat{K}_{\hat{k}, \hat{k}} = \frac{1}{d}\sum_{j = 1}^{d}
s^{(\gamma)}(k_{_j})$. The $i^{\thh}$ element of the vectors containing the
dependence on $\nvec$, and $\novec$ is given, respectively, by
${\mvec{r}_{\nvec_i}} = \prod_{j = 1}^{d} g_{r}^{(\gamma_j)}(n_{i_j},
k_{_j})$ and
${\mvec{\ell}_{{\novec}_i}} = \prod_{j = 1}^{d} g_{\ell}^{(\gamma_j)}(n_{0_{i_j}}, k_{_j})$. The remainder of the procedure is
identical to the 1D case outlined previously.

\subsection{Eigenvectors and eigenvalues of transition matrix of the one
  dimensional Lattice random walk}
\label{sec:eigvecs_eigvals_props}
Since the high dimensional eigenvectors and eigenvalues are composed
of the 1D case, repeat here the quantities of interest diffusive case
as an example. The Master equation governing a random walk on a finite
lattice with $N$ distinct sites $(1 \leq n \leq N)$ is written as
\begin{equation}
  \label{}
\propt(n, t + t) = \sum_{m = 1}^N \mmat{A}_{n, m}\propt(m, t),
\end{equation}
where  $\mmat{A}$ is the transition matrix. The matricial form allows
one to write easily the propagator as 
\begin{equation}
  \label{eq:one_d_prop_mat}
\proptt{n}{n_0} = \sum_{k = 1}^{N} g_{r}^{(\gamma)}(n, k)  g_{\ell}^{(\gamma)}(n_0, k) s^{(k)}(k)^t,
\end{equation}
where $g_{r}^{(\gamma)}(n, k)$ and $g_{\ell}^{(\gamma)}(n, k)$ are the
$n^{\mathrm{th}}$ component of, respectively, the $k^{\mathrm{th}}$
right and left eigenvectors of the transition matrix and
$s^{(\gamma)}(k)$ is the $k^{\mathrm{th}}$ eigenvalue. The right
eigenvectors are given by
\begin{equation}
  \label{eq:gr}
g_{r}^{(\gamma)}(n, k)
 = 
\lc
\begin{matrix}
  \frac{a_k}{\sqrt{N}}\cos{\ls \lb n - \frac{1}{2} \rb \frac{(k - 1) \pi}{N} \rs} & \gamma = r, \\
  \frac{1}{\sqrt{N}}\exp{\ls \frac{2 \pi n i \lb k-1 \rb}{N} \rs} & \gamma = p, \\
  \sqrt{\frac{2}{N}}\sin{\ls \lb \frac{n - 1}{N - 1} \rb k \pi \rs} & \gamma = a, \\
  \frac{2}{\sqrt{2N - 1}}\cos{\ls \lb n -  \frac{1}{2} \rb \frac{2k - 1}{2 N - 1} \pi \rs} & \gamma = m,
\end{matrix}
\rd
\end{equation}
with $a_k = 2$ for $k = 1$ and $a_k = 1$ for all other values of $k$,
while the left eigenvectors, $g_{\ell}^{(\gamma)}(n, k)$ are identical
to \cref{eq:gr} for all cases except the periodic boundary condition
$\gamma = p$, where instead it is given by
$g_{\ell}^{(p)}(n, k) = 1 / g_{r}^{(p)}(n, k)$, finally the eigenvalues
can be found in Eq.~(4) of Ref.~\cite{giuggioli2020}. For periodic
($\gamma = p$) and reflecting domains ($\gamma = r$) with $N$ distinct
sites we have $k \in [1, N]$ eigenvalues and eigenvectors, for
absorbing boundary conditions ($\gamma = a$) with absorbing sites at
$n = 1$ and $n = N$ we have $k \in [1, N-2]$ eigenvalues, lastly for
mixed boundary condition ($\gamma = m$) with a reflecting end at
$n = 1$ and an absorbing one at $n = N$ we have $k \in [1, N-1]$
eigenvalues.

\bibliography{library1.bib,library2.bib,library3.bib} 
\makeatletter\@input{xx.tex}\makeatother